\documentclass[%
 aip,
 amsmath,amssymb,
 reprint,%
]{revtex4-1}

\usepackage{graphicx}
\usepackage{dcolumn}
\usepackage{bm}

\usepackage[utf8]{inputenc}
\usepackage[T1]{fontenc}
\usepackage{mathptmx}
\usepackage{etoolbox}

\usepackage{natbib}
\usepackage{tikz,pgfplots}
\usepackage[caption=false]{subfig}
\usepackage{multirow}
\usepackage{float}

\makeatletter
\def\@email#1#2{%
 \endgroup
 \patchcmd{\titleblock@produce}
  {\frontmatter@RRAPformat}
  {\frontmatter@RRAPformat{\produce@RRAP{*#1\href{mailto:#2}{#2}}}\frontmatter@RRAPformat}
  {}{}
}%
\makeatother
\begin{document}

\preprint{AIP/123-QED}

\title[Ponderomotive-expulsion: toward creating an electron-free volume]{Ponderomotive-expulsion: toward creating an electron-free volume}

\author{Smrithan Ravichandran}
\affiliation{%
 Institute for Physical Science and Technology, University of Maryland, College Park, Maryland 20742, USA}%
\affiliation{Joint Quantum Institute, University of Maryland, College Park, Maryland 20742, USA}

\author{Teresa Cebriano}
\affiliation{Centro de L\'aseres Pulsados, 37185 Villamayor, Salamanca, Spain}

\author{Jos\'e Luis Henares}
\affiliation{Centro de L\'aseres Pulsados, 37185 Villamayor, Salamanca, Spain}

\author{Cruz Mendez}
\affiliation{Centro de L\'aseres Pulsados, 37185 Villamayor, Salamanca, Spain}

\author{Jos\'e Antonio P\'erez-Hern\'andez}
\affiliation{Centro de L\'aseres Pulsados, 37185 Villamayor, Salamanca, Spain}

\author{Luis Roso}
\affiliation{Departamento de F\'isica Aplicada, Universidad de Salamanca, 37008 Salamanca, Spain}

\author{Robert Fedosejevs}
\affiliation{Electrical and Computer Engineering, University of Alberta, Edmonton, Alberta T6G 2V4, Canada}

\author{Wendell T. Hill, III}
\email{wth@umd.edu}
\affiliation{%
 Institute for Physical Science and Technology, University of Maryland, College Park, Maryland 20742, USA}%
\affiliation{Joint Quantum Institute, University of Maryland, College Park, Maryland 20742, USA}
\affiliation{Department of Physics, University of Maryland, College Park, Maryland 20742, USA}


\begin{abstract}

We describe a demonstration of a prototype approach to clear the laser focal volume of free electrons and disable their atomic and molecular sources. Employing two temporally separated, copropagating pulses, we exploited a pump-probe setup in our experiment. The pump ionized a low-density gas and expelled free and nascent electrons from its focal volume. The probe, traversing the same focal volume, expelled any remaining free and probe-induced nascent electrons. We gauged the effectiveness of the approach by capturing the spatial distribution of ejected electrons with image plates while we varied the relative intensity and time delay between the pump and probe. When we injected the pump 300 fs before the probe, we found the electron spatial distribution significantly altered and the yield suppressed, proving ponderomotive expulsion works. However, the yield was enhanced when we set the temporal spacing between the pump and probe to 150 fs. Simulations show the enhancement is due to Airy rings of the focused pump expelling electrons inward toward the propagation axis. Our results show that the complete removal of focal-volume electrons was inhibited by spatial overlap fluctuations and the stronger probe generating ionization outside the cleared-volume of the pump. We discuss ways to mitigate these impediments and propose alternate two-beam arrangements to achieve more efficient focal-volume clearing.

\end{abstract}

\maketitle

\section{Introduction}\label{sec:Introduction_section}

Recent and ongoing developments of state-of-the-art tools such as petawatt lasers \cite{peng_overview_2021,hernandez-gomez_vulcan_2010,papadopoulos_apollon_2024} and X-ray Free Electron Lasers (XFEL) \cite{galayda_lcls-ii_2018,raubenheimer_lcls-ii-he_2018,decking_commissioning_2017,nolle_commissioning_2018} may soon enable an unprecedented exploration of the quantum vacuum (QVAC) \cite{heinzl_observation_2006,lundstrom_using_2006,bell_possibility_2008,tommasini_precision_2009,tommasini_precision_2009,tommasini_light_2010,krajewska_bethe-heitler_2013,zhu_dense_2016,king_measuring_2016}. Several experiments have been proposed to study the nearly century-old QVAC theory \cite{gies_photon-photon_2018,king_three-pulse_2018,mosman_vacuum_2021,gies_all-optical_2022,roso_towards_2022,ahmadiniaz_detection_2023,berezin_calculation_2023,berezin_analytical_2024,formanek_signatures_2024,noauthor_aps_nodate,noauthor_aps_nodate-1,ahmadiniaz_towards_2025} in novel regimes enabled by the use of petawatt lasers and XFELs. However, the presence of real particles, such as electrons, in the interaction volume can impede the detection of the extremely weak light-light scattering signal associated with nonlinear terms in the QVAC Lagrangian; ordinary electron-light scattering and seeded electron-positron pair production\cite{grismayer_seeded_2017,sampath_towards_2018,mironov_onset_2021,gonoskov_charged_2022} are two processes that need to be suppressed. Therefore, creating a focal volume devoid of free electrons and incapable of producing free electrons prior to arrival of the main pulse(s) is a critical need to enable precision measurements of the QVAC. While ultra-high vacuum pressures of $\sim 10^{-14}\ \mathrm{mbar}$ may be achievable in vacuum chambers with specially-designed vacuum technologies \cite{benvenuti_obtention_1993}, they are not feasible in chambers that are regularly exposed to the atmosphere. The presence of water vapor in the atmosphere prevents chambers that undergo frequent ventilation in high-power laser facilities from reaching pressures $\lesssim 10^{-7} \ \mathrm{mbar}$. This corresponds to a particle density of $\sim 3\times10^9 /\mathrm{cm^3}$. One proposed technique to reduce the density of atoms and molecules in the focal volume is to use a laser pulse to ionize and a static electric field to subsequently expel the ions\cite{yu_definite_2025}. For this technique, delay times of several nanoseconds are required and electrodes must be placed close to the focal volume. Another approach is to just expel all electrons from the interaction region as the intense fields interact primarily with electrons. Creating an electron-free void in the sub-nanosecond timescale can therefore be highly beneficial. One approach to expel electrons from rarefied gas particles, mostly light gases\cite{redhead_hydrogen_2003} like $\mathrm{H_2}$, is to allow a fraction of the laser energy to pass through the focus, known as a ``pre-pulse" or ``pedestal", before the main pulse. It is expected that the pre-pulse or pedestal of sufficient intensity will ionize and clear resultant electrons via ponderomotive expulsion. While there is evidence that electrons are swept out of the focus by the ponderomotive force, there is little information on how strong a pre-pulse is needed. 



In this manuscript, we report on experimental evidence of ponderomotive electron expulsion from the focal volume of petawatt-class lasers using two laser pulses separated by a few hundred femtoseconds in copropagating geometry. It is well known that at relativistic laser intensities, electrons liberated from rarefied gas in the focal volume are ejected from the focus at an acute angle\cite{moore_observation_1995}, $\theta$, relative to the laser axis. Under paraxial conditions, i.e., large f-numbers such as the f/10 focus used here, these electrons are ejected down to some minimum angle, $\theta_c$, which is characteristic of the peak intensity\cite{ravichandran_imaging_2023,longman_toward_2023}, scaling as $\tan\theta_c \propto 1/a_0$. Here, $a_0$ is the normalized vector potential, $a_0 \approx 0.855 \lambda_0 \mathrm{(\mu m)} \sqrt{I_0\mathrm{(W/cm^2)/10^{18}}}$, with $\lambda_0$ and $I_0$ being the central pulse wavelength and peak intensity, respectively. We can thus directly probe the concentration of electrons available within the focal volume by measuring the spatial distribution of the ejected electrons. Further, by deploying two copropagating laser pulses separated in time, we can observe the effect of allowing a pulse to pass through the focal volume before a trailing pulse. Through this, we can investigate the clearing mechanism of a pre-pulse and gain insights toward requirements for efficient clearing. 

\begin{figure}[htp]
    \centering
    \begin{minipage}{1\columnwidth}
        \subfloat[\label{subfig:Strong_leading_schematic}]{\includegraphics[height=3cm,trim={0cm 0cm 0cm 0cm},clip]{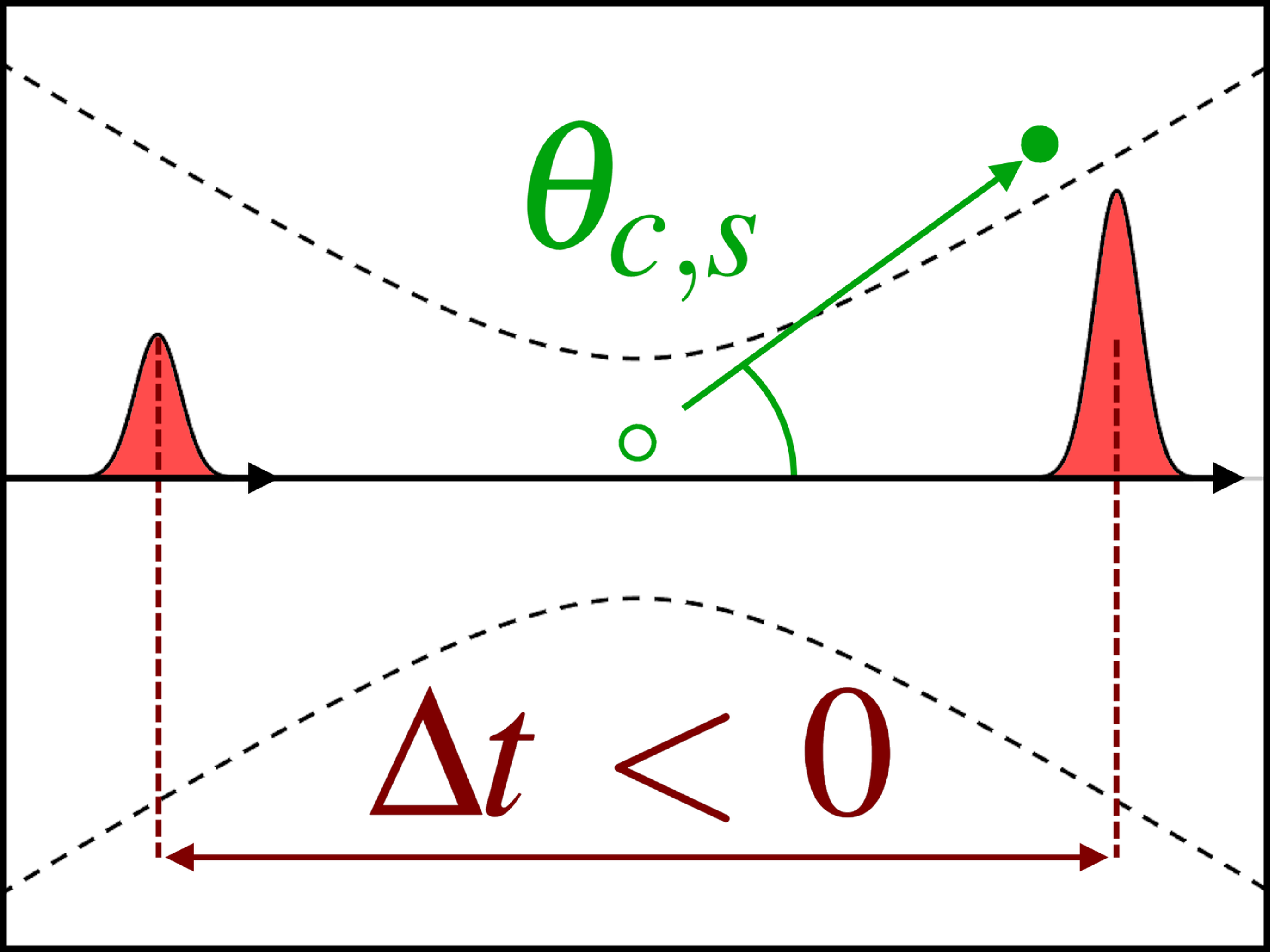}}\hspace{1mm}
        \subfloat[\label{subfig:Weak_leading_schematic}]{\includegraphics[height=3cm,trim={0cm 0cm 0cm 0cm},clip]{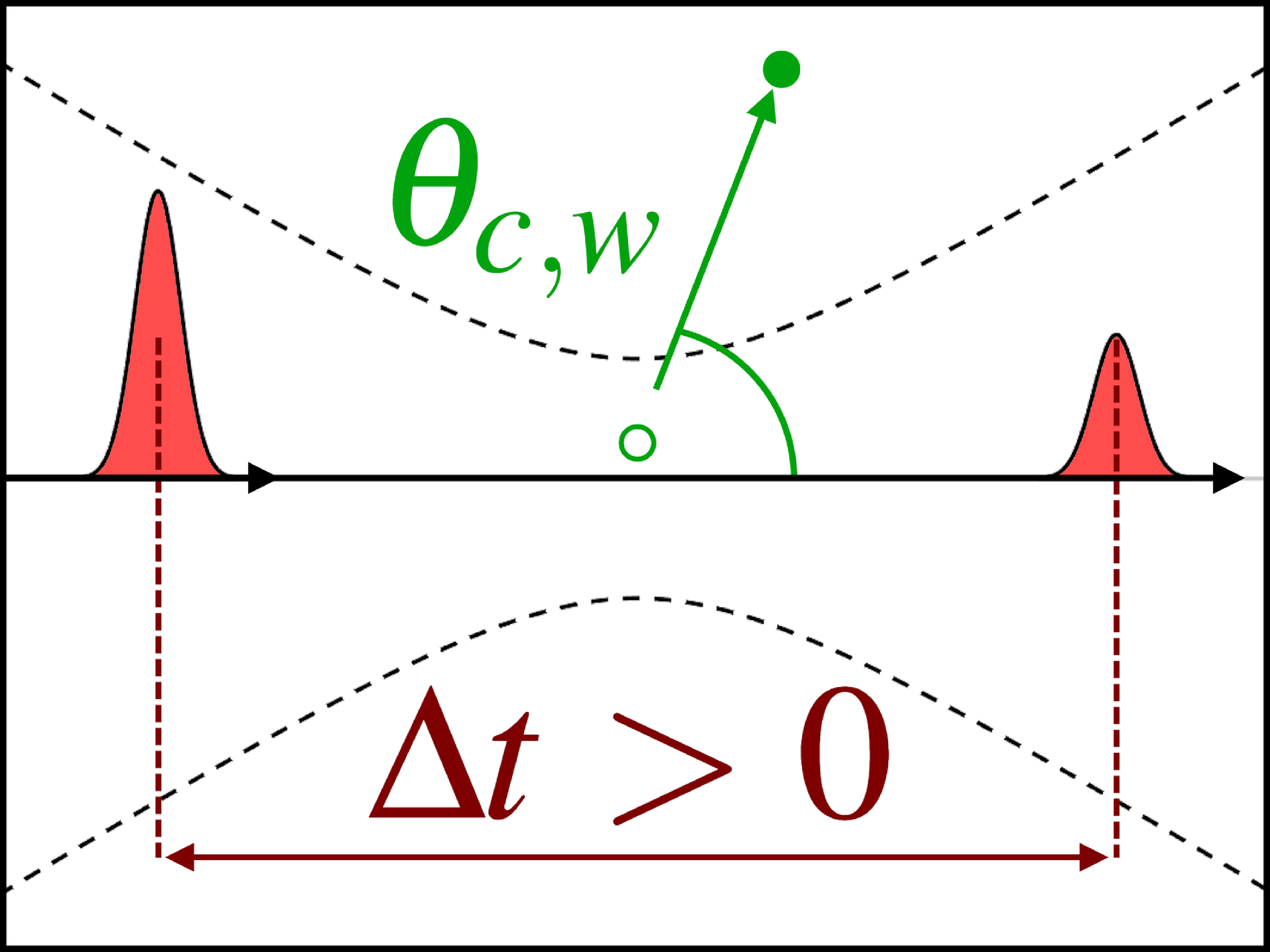}} 
    \end{minipage}\\
    \vspace{3mm}
    \hspace{-7mm}
    \begin{minipage}{1\columnwidth}
        {\includegraphics[height=1.30cm,trim={-0.17cm 8.7cm 0.7cm 1.85cm},clip]{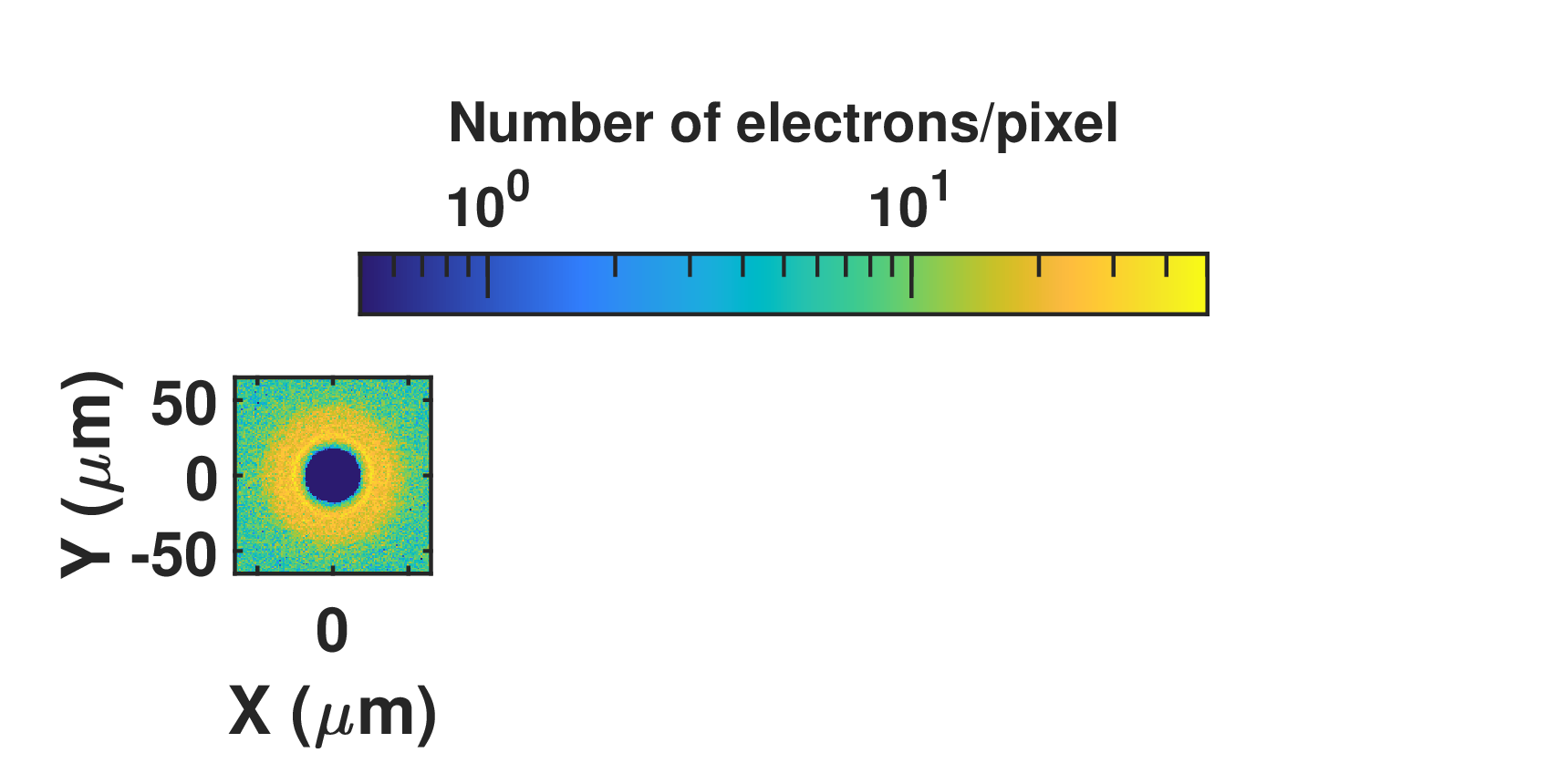}}
    \end{minipage}\\
    \vspace{1.5mm}
    \begin{minipage}{.75\columnwidth}
        \hspace{2.3mm}\subfloat[\label{subfig:Sim_300fs_s}]{\hspace{-1cm}\includegraphics[height=3.42cm,trim={0cm 1.75cm 7.1cm 1.7cm},clip]{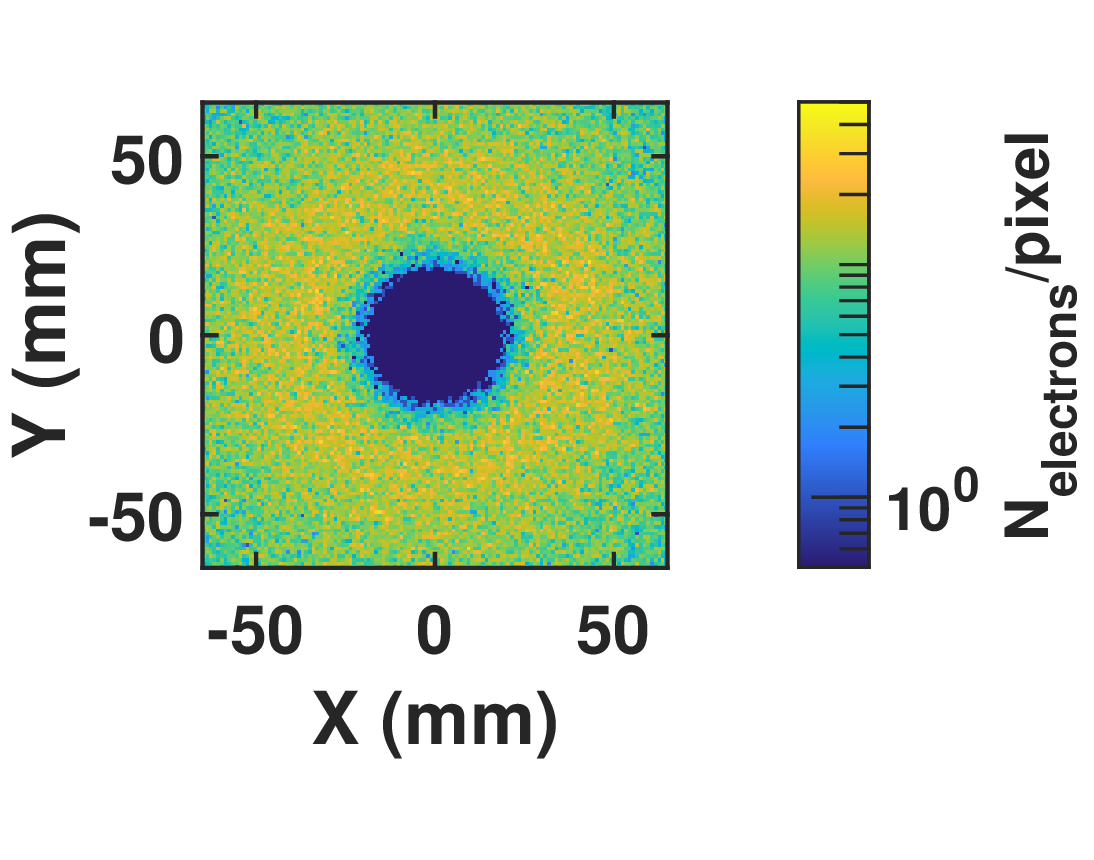}}
        \hspace{1.2mm}\subfloat[\label{subfig:Sim_300fs_w}]{\hspace{0cm}\includegraphics[height=3.42cm,trim={3.2cm 1.75cm 7.1cm 1.7cm},clip]{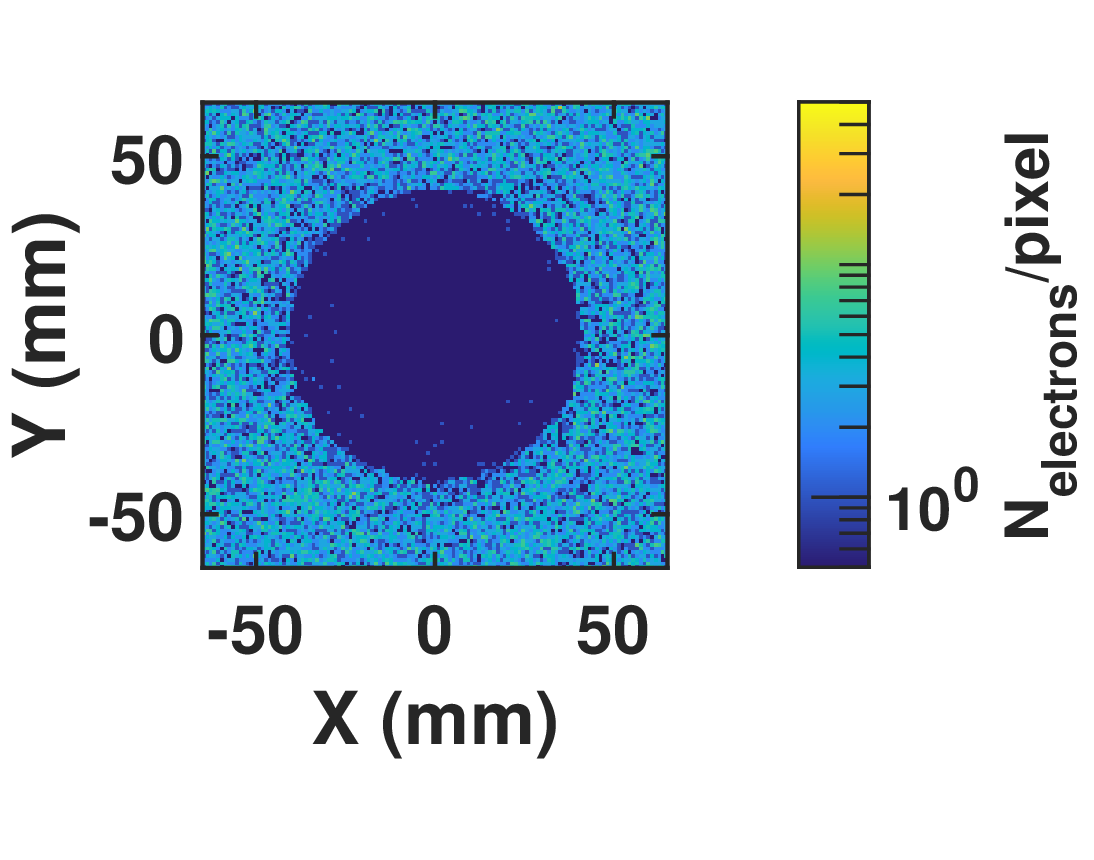}} 
    \end{minipage}\\
    \vspace{-3.2mm}
    \subfloat[\label{subfig:Sim_300fs_lineouts}]{\includegraphics[width=0.8\linewidth]{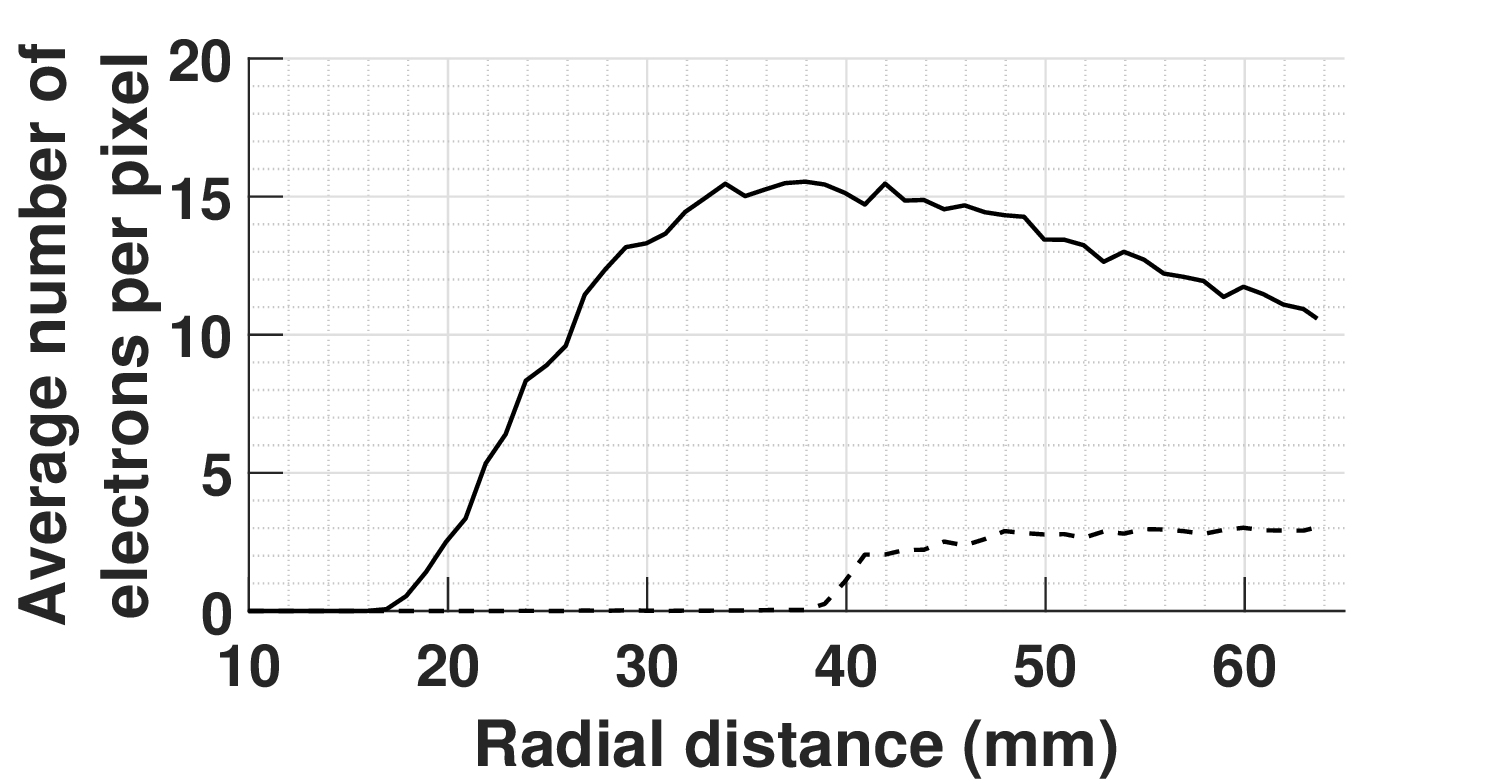}}\\
\caption{A cartoon representing the minimum angle of ejection for the (a) strong pulse, $\theta_{c,s}$ and (b) weak pulse, $\theta_{c,w}$. Spatial distribution of ejected electrons simulated when the (c) strong pulse and (d) weak pulse leads the other by $300\ \mathrm{fs}$ with perfect overlap of focal volumes. Distributions simulated 30 mm after the focus on a plane perpendicular to the laser axis passing through the origin. The color axis denotes the number of electrons in each $1\ \mathrm{mm} \times 1\ \mathrm{mm}$ pixel across the plane. The corresponding average number of electrons for varying distance from the laser axis is shown by the solid and dashed lines respectively in (e). Simulation details in Appendix \ref{sec:appendix_simulations}.}
\label{fig:300fs_sim_ideal_overlap}
\end{figure}

The first laser pulse liberates and accelerates electrons from rarefied gas ($\sim$ few $\times 10^{-4}$ mbar, $\gtrapprox90\%$ He) in the focal volume. The second laser pulse probes the focal volume a few hundred femtoseconds later. This allows us to detect electrons that may have remained in the focal volume after the first pulse. Here, we use two pulses named strong and weak, separated by a time delay $\Delta t$. We define $\Delta t<0$ ($\Delta t > 0$) when the strong (weak) pulse leads the weak (strong) as shown in Fig.~\ref{subfig:Strong_leading_schematic} (Fig.~\ref{subfig:Weak_leading_schematic}). The discernibility of electrons ejected from the focal volume in either case is based on the notable difference in peak intensities of the strong and weak pulses, $I_{0s}$ and $I_{0w}$, respectively, i.e., $I_{0s}>I_{0w}$. Here, $I_{0w}\gg10^{16}\ \mathrm{W/cm^2}\sim$ the intensity required to doubly ionize He, implying that both laser pulses can deplete electrons from He over a large volume ($\rho \gtrsim 3w_0$, $|z|\gtrsim500z_R$, where $\rho$ is the radial distance from the laser axis, $z$ is the distance from the focal plane along the laser axis and $z_R$ is the Rayleigh length). When the strong pulse leads the weak, we expect that electrons liberated by the strong pulse are ejected at some $\theta\geq\theta_{c,s}$. When the weak pulse leads the strong, however, we expect that the weak pulse expels most electrons from the focal volume over $\theta\geq \theta_{c,w}$, leaving behind a paucity of electrons to be ejected by the strong pulse. As $\tan \theta _c \propto 1/\sqrt{I_0}$ and $I_{0w}<I_{0s}$, we note that $\theta_{c,w}>\theta_{c,s}$ as depicted by the cartoon in Figs.~\ref{subfig:Strong_leading_schematic} and \ref{subfig:Weak_leading_schematic}. Therefore, we expect the yield of electrons for $\theta_{c,s}<\theta<\theta_{c,w}$ to be lower when the weak pulse leads the strong pulse, relative to when the strong pulse leads the weak pulse, or when the weak pulse is absent.


We demonstrate the design of the experiment by simulating the interaction of electrons generated from He with two copropagating laser pulses having Gaussian beam profiles. We provide details of our simulations in Appendix \ref{sec:appendix_simulations}. We show the spatial distribution of electrons simulated on a plane perpendicular to the laser axis ($\hat{z}$), 30 mm after the laser focus in Fig.~\ref{subfig:Sim_300fs_s} (Fig.~\ref{subfig:Sim_300fs_w}) when the strong (weak) pulse leads the weak (strong) pulse by 300 fs. We characterize each two-dimensional image of the electron distribution by $N_e(\Delta t,\rho)$, the average number of electrons deposited per pixel vs radial distance, $\rho=\sqrt{x^2+y^2}$, from the laser axis in Fig.~\ref{subfig:Sim_300fs_lineouts}. Here,

\begin{equation} \label{eqn:line_profile}
    N_e(\Delta t,\rho) = \frac{1}{2\pi}\int_{\phi=0}^{\phi=2\pi} n_e(\Delta t,\rho,\phi) \,d\phi
\end{equation}


\noindent where $\phi=\tan^{-1}(y/x)$ and $n_e(\Delta t,\rho,\phi)$ is the number of electrons incident on a pixel at some $(\rho,\phi)$ on the image plate. We see that the electron yield is significantly reduced when the weak pulse leads the strong pulse. Therefore, our goal is to measure the spatial distribution of ejected electrons and compare the resulting $\phi$-averaged line profiles for $\Delta t >0$ (weak pulse leading) and $\Delta t<0$ (strong pulse leading). A reduction in electron yield for varying $\theta$ when $\Delta t >0$ relative to $\Delta t<0$ will provide evidence of ponderomotive expulsion from the focal volume. 

The remainder of this manuscript is organized as follows. In Sec.~\ref{sec:methods_section}, we present our experimental setup and measurement of the spatial distribution of ejected electrons. In Sec.~\ref{sec:discussion_section}, we compare our measurements and simulations performed with $\Delta t=\pm 300$ and $\pm150 \ \mathrm{fs}$, and highlight the role of realistic features that exist in the experiment. In addition, we suggest ways to observe more efficient clearing and a way forward for future experiments, and provide concluding remarks in Sec.~\ref{sec:conclusion_section}.

\section{Methods}\label{sec:methods_section}

\begin{figure}[htp]
    \centering
    \hspace{1mm}\subfloat[\label{subfig:Exp_schematic_subfig}]{\includegraphics[width=0.7\linewidth]{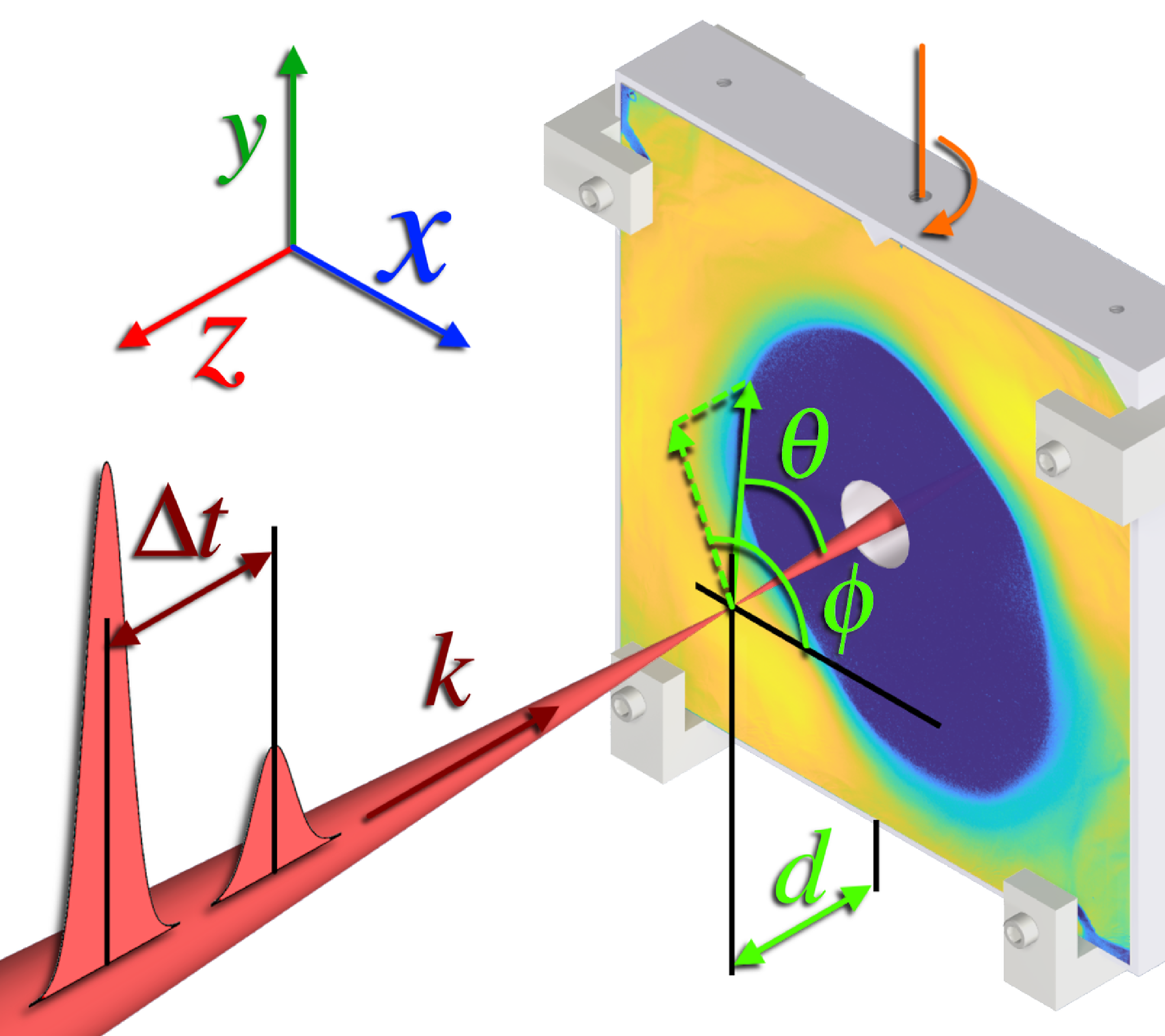}}
\begin{minipage}[s]{.45\columnwidth}
    \vspace{-4.54cm}\hspace{-5mm}\subfloat[\label{subfig:Focal_spot_strong}]{\hspace{-0.5cm}\includegraphics[width=2.25cm,trim={2.5cm 5.25cm 5cm 1.8cm},clip]{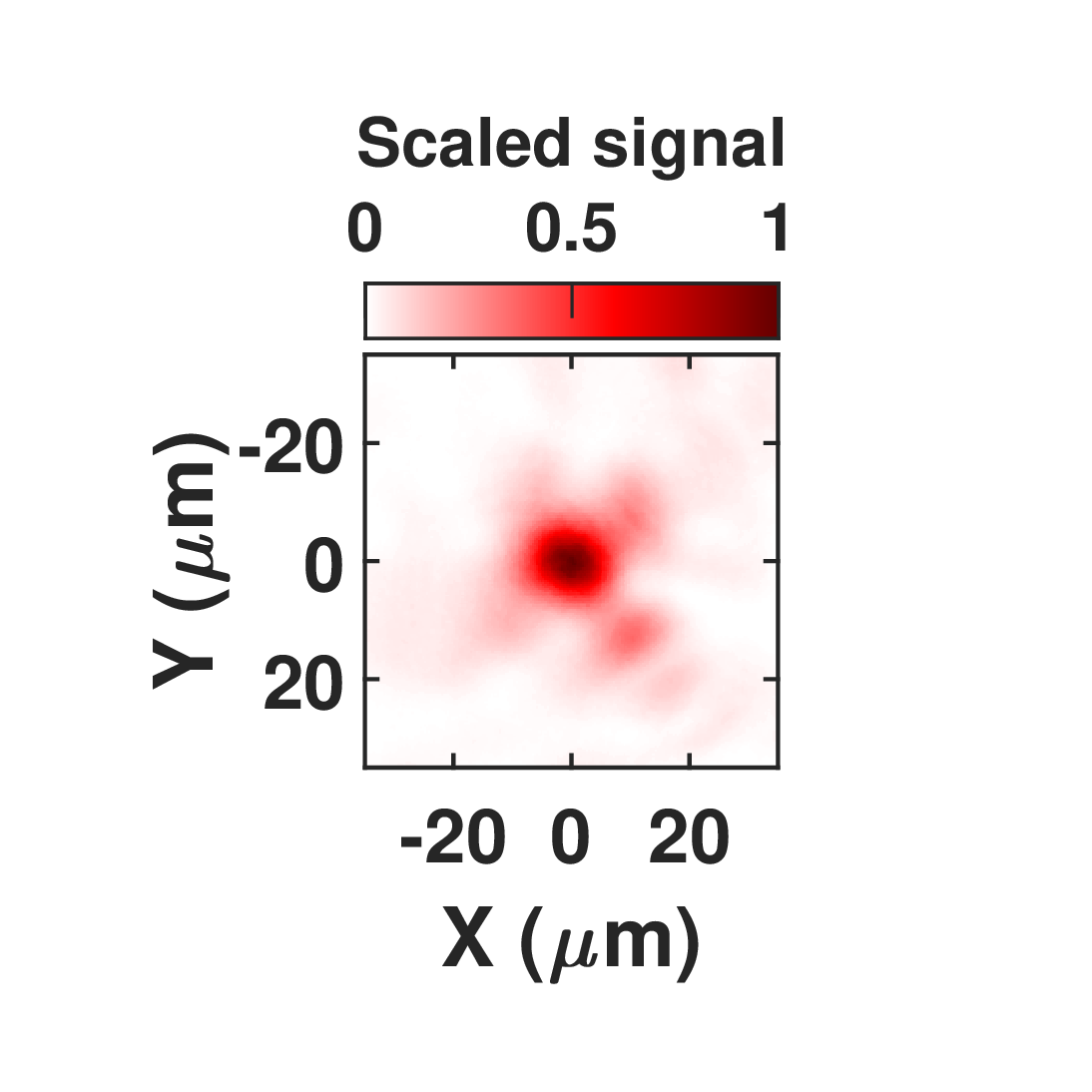}}
    \vfil \vspace{1mm}
    \hspace{-5mm}\subfloat[\label{subfig:Focal_spot_weak}]{\hspace{-0.5cm}\includegraphics[width=2.25cm,trim={2.5cm 1.75cm 5cm 6cm},clip]{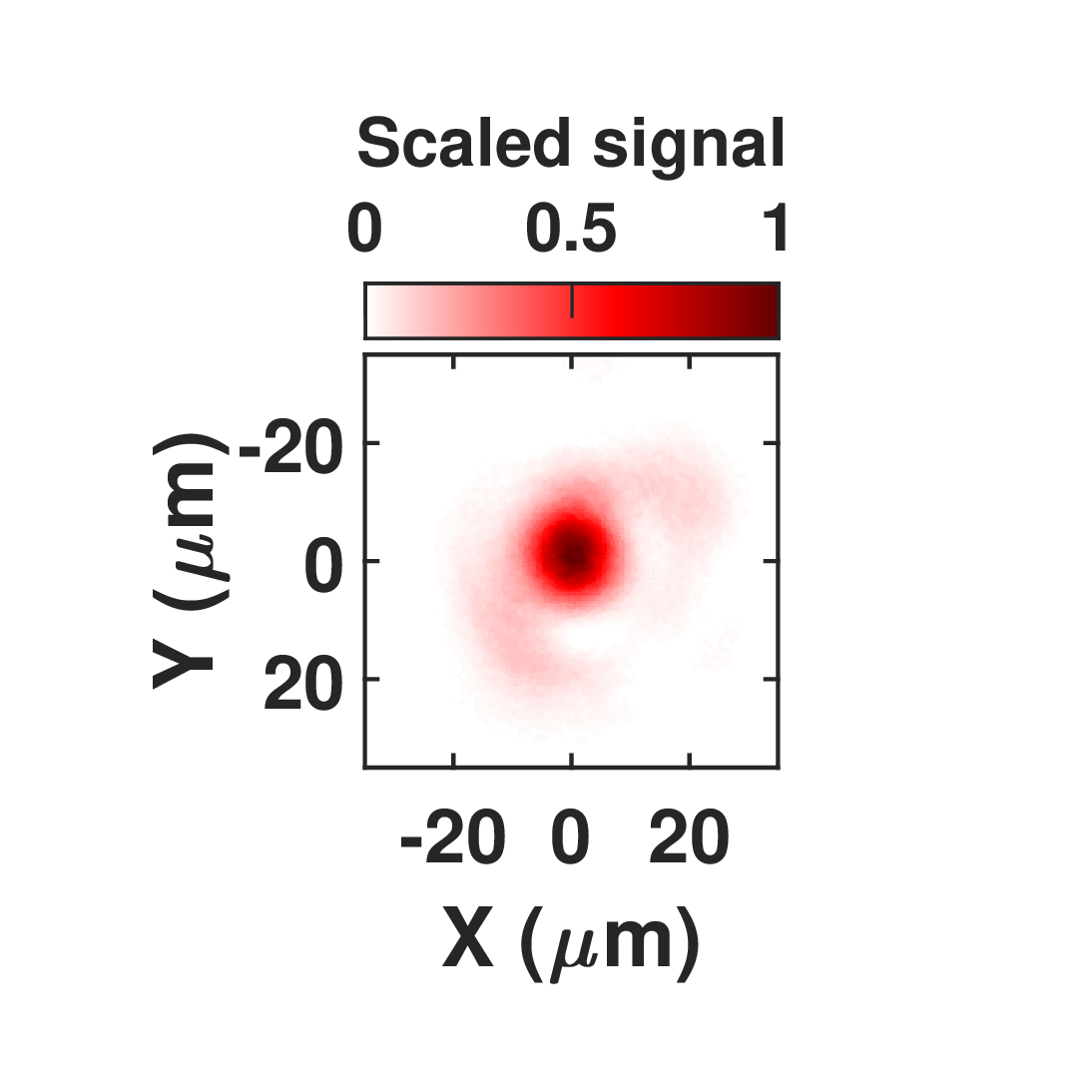}}
\end{minipage}
\caption{Schematic of the experimental setup (a) and typical isolated focal-spot images of the strong (b) and weak (c) pulses. The laser pulses propagate along $-\hat{z}$, polarized predominantly along $\hat{x}$ and separated in time by $\pm|\Delta t|$. A pair of image plates are loaded onto the image-plate holder that can rotate about the vertical axis indicated in orange.}
\label{fig:Exp_setup}
\end{figure}

We used the VEGA-3 laser beamline at the Centro de L\'aseres Pulsados (CLPU)\cite{noauthor_clpu_nodate} in Spain for our experiment. The nominal pulse energy was $\sim25$ J. We split the fully amplified pulse before the pulse compressor using a Mach-Zehnder interferometer to create two copropagating pulses with $\sim$10 J and 2.5 J. The beam transport system delivered $\sim87\%$ of the energy to the focal plane. The nominal beam diameter at the Mach-Zehnder interferometer was $\sim$ 10 cm. We used an apodizer of diameter 6 cm on the lower-energy beam arm to increase the beam waist and improve the spatial quality of the weak pulse at the focus. We verified spatial overlap of the two foci prior to each measurement by imaging a series of 20 shots at low power after setting the desired time delay. We measured the pulse duration, $\tau \sim$ 40 fs, by diverting a small portion of the beam to a second-order intensity autocorrelator outside the target chamber. We also used the autocorrelation measurement, which in the presence of two pulses contains a pair of secondary peaks spaced $\pm|\Delta t|$ away from the main peak, to verify that the set delay is delivered to the target chamber. We inferred the nominal peak intensities in the experiment, $I_{0s}\sim 6.6\times 10^{19}$ ($a_{0s} \approx 5.6$) and $I_{0w}\sim 7\times 10^{18} \ \mathrm{W/cm^2}$ ($a_{0w} = 1.8$) for the strong and weak pulse, respectively, by distributing the peak power ($U/\tau$) across the focal spot image as detailed in the Appendix of C.Z.~He et al.\cite{he_towards_2019} 

We used He gas to mimic the gaseous composition of a high-vacuum environment where $\mathrm{H_2}$ is most prevalent, as He is fully ionized at non-relativistic intensities $\sim 10^{16} \ \mathrm{W/cm^2}$, which is sufficient to split $\mathrm{H_2}$ into two protons and two electrons. We backfilled the chamber to $\approx 2.75 \times 10^{-4}$ mbar of He from a base pressure $\sim$ a few $\times \ 10^{-5}$ mbar. We used a residual gas analyzer to ensure that the partial pressure of He was $\gtrapprox$ 90\% during operation, with other gases such as $\mathrm{N_2, H_2O, O_2}$ making up the rest.

\begin{figure}[htp]
    \centering
    \begin{minipage}{1\columnwidth}
        \hspace{-6.3mm}{\includegraphics[height=1.42cm,trim={-0.45cm 24.1cm 4.8cm 0.37cm},clip]{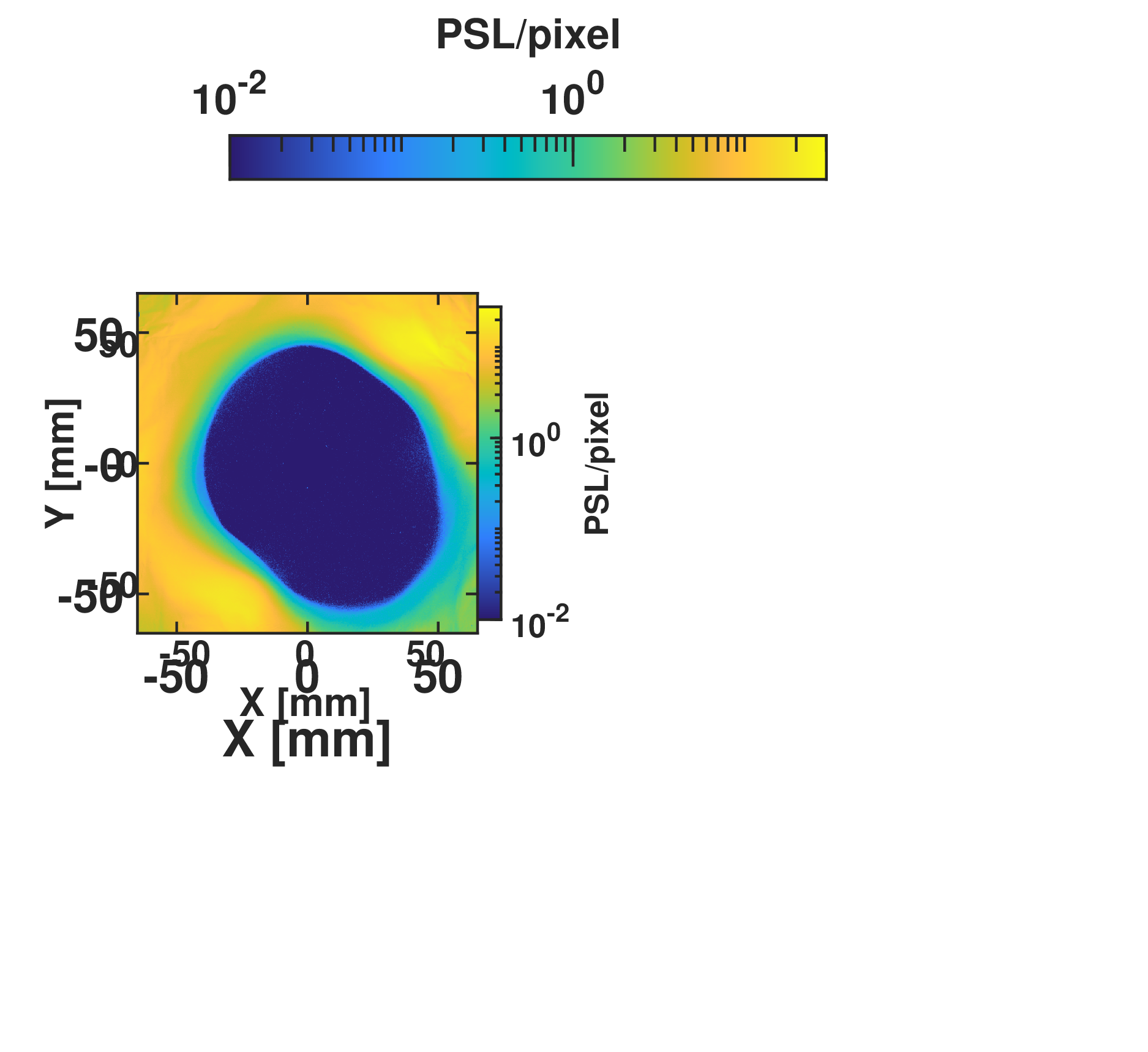}}
    \end{minipage}\\
    \hspace{0mm}
    \begin{minipage}{.75\columnwidth}
        \vspace{-2.5mm}
        \hspace{0cm}\subfloat[\label{subfig:Exp_sample_300fs_s}]{\hspace{-0.8cm}\includegraphics[height=3.44cm,trim={19.4cm 2.0cm 8.5cm 1.5cm},clip]{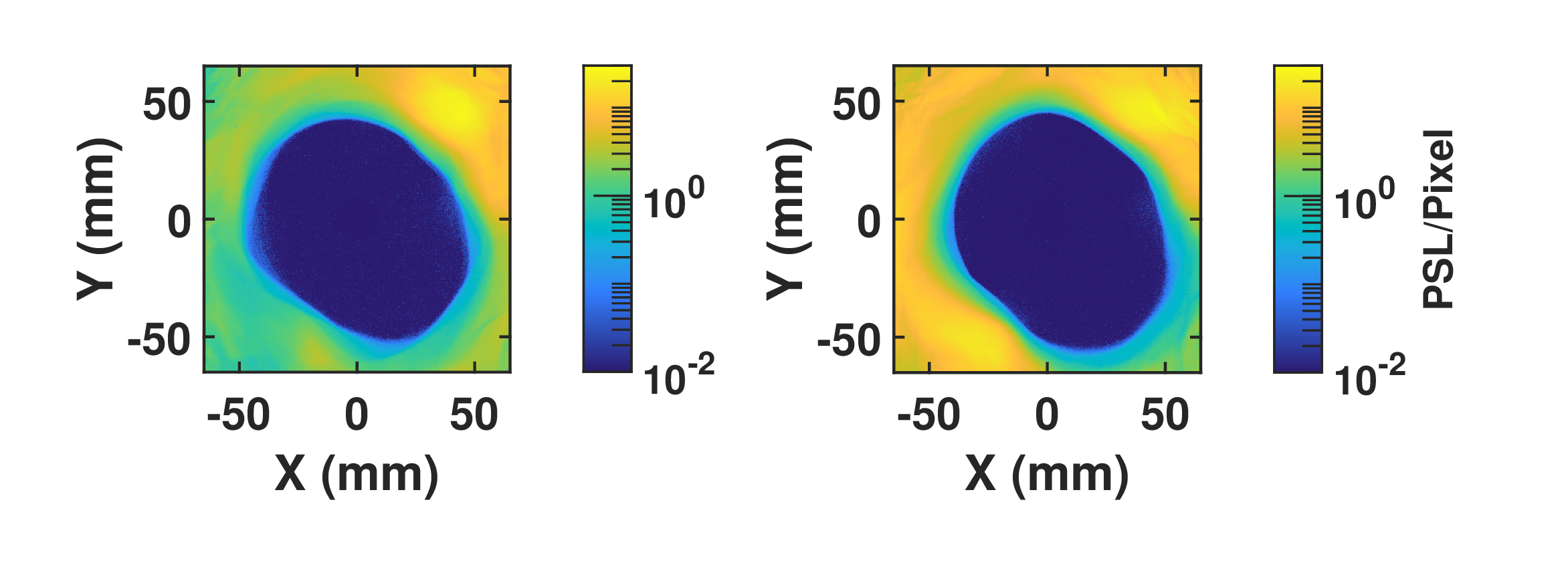}}
        \hspace{-1mm}\subfloat[\label{subfig:Exp_sample_300fs_w}]{\hspace{0cm}\includegraphics[height=3.44cm,trim={4.9cm 2.0cm 26.5cm 1.5cm},clip]{EPS_Exp_sample_300fs.eps}} 
    \end{minipage}\\
    \vspace{-3mm}
    \subfloat[\label{subfig:Exp_300fs_lineouts}]{\includegraphics[width=0.8\linewidth]{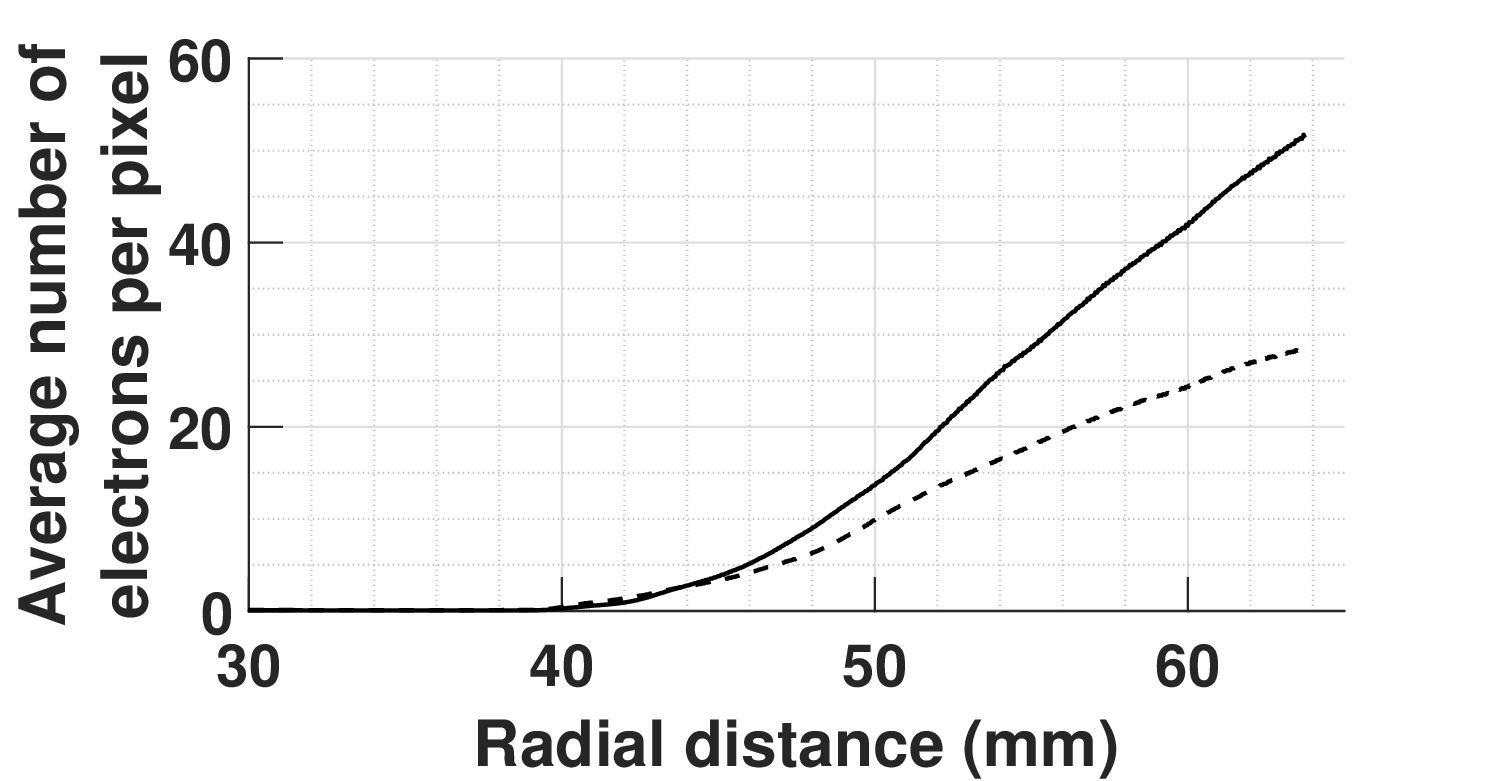}}\\
\caption{Spatial distribution of ejected electrons measured when the (a) strong pulse and (b) weak pulse leads the other by $300\ \mathrm{fs}$. Distributions measured 30 mm after the focus on a plane perpendicular to the laser axis passing through the origin. The color axis denotes the number of electrons in each $50\ \mathrm{\mu m} \times 50\ \mathrm{\mu m}$ pixel across the plane. The corresponding average number of electrons for varying distance from the laser axis is shown by the solid and dashed lines respectively in (c). Measurement details in Sec.~\ref{sec:methods_section}.}
\label{fig:300fs_experiment}
\end{figure}

We used a Fujifilm BAS-MS image plate placed at a distance $d =$ 30 mm after, and facing, the laser focus to detect ejected electrons on a plane perpendicular to the laser axis, $\vec{k}$, as shown in Fig.~\ref{fig:Exp_setup}. The size of the image plate was $\sim 148 \ \mathrm{mm} \times 148 \ \mathrm{mm}$ with a 20-mm diameter hole cut in the center to allow the laser to pass through. We covered the image plate with an aluminum foil (thickness 12 $\mathrm{\mu m}$) to make it insensitive to laser light. We used a holder designed to accommodate two image plates at once such that one faces the laser focus during measurement while the other faces away. We made the holder using PEEK plastic of sufficient thickness, 25.4 mm, to ensure that the image plate facing away was not exposed to electrons from the focus during measurement. Each measurement sequence consisted of 50 shots, each shot involving the passing of both pulses. After completing one measurement, we rotated the holder by $180^{\circ}$ about an axis centered within the thickness of the holder intersecting the laser axis to interchange the image plates for the second measurement as shown in Fig.~\ref{fig:Exp_setup}. Then, we vented the vacuum chamber to scan the image plates. We made one measurement with $\Delta t = |\Delta t|\ (>0)$ and another with $\Delta t = -|\Delta t|\ (<0)$ within one vacuum cycle to minimize the role of systematic uncertainties, such as variations in pulse duration or focal spot quality, which can affect the peak intensity. Image plate measurements from a single vacuum cycle corresponding to $\Delta t = \pm300 \ \mathrm{fs}$ are shown in Figs.~\ref{subfig:Exp_sample_300fs_s} and \ref{subfig:Exp_sample_300fs_w} in units of phosphostimulated luminescence (PSL) \cite{williams_calibration_2014} per pixel (size $50 \ \mathrm{\mu m} \times 50 \ \mathrm{\mu m}$), which is directly proportional to the energy deposited in the pixel. 

To obtain a $\phi$-averaged line profile for our measurements, we first convert the PSL value per pixel to an estimate of the number of electrons per pixel by taking into account the electron energy as a function of angle and response functions of the filters and image plates, as described in Appendix \ref{sec:appendix_PSL_to_electrons}. We then calculate $N_e(\Delta t,\rho)$ from $n_e(\Delta t,\rho,\phi)$ using Eq.~\ref{eqn:line_profile}. Figure \ref{subfig:Exp_300fs_lineouts} shows $N_e(-300 \ \mathrm{fs},\rho)$ and $N_e(300\ \mathrm{fs},\rho)$ for measurements in Figs.~\ref{subfig:Exp_sample_300fs_s} and \ref{subfig:Exp_sample_300fs_w}, respectively.

\section{Discussion} \label{sec:discussion_section}

Electrons are cleared from the focal volume by the weak pulse if the $\phi$-averaged line profile $N_{e}(\Delta t,\rho)$ is reduced when the weak pulse leads relative to when the strong pulse leads. This is shown both by simulation and experiment in Figs.~\ref{subfig:Sim_300fs_lineouts} and \ref{subfig:Exp_300fs_lineouts}, respectively, implying that clearing is observed at $|\Delta t|= 300\ \mathrm{fs}$. However, we see two features in Fig.~\ref{subfig:Sim_300fs_lineouts} that do not accurately reflect the experiment. Namely, (i) $N_e$ rises as early as $\rho \approx 17\ \mathrm{mm}$ in Fig.~\ref{subfig:Sim_300fs_lineouts} while it rises at $\rho\approx 40\ \mathrm{mm}$ in Fig.~\ref{subfig:Exp_300fs_lineouts}, and (ii) $N_e$ appears distinctly different in Fig.~\ref{subfig:Sim_300fs_lineouts} depending on which pulse leads i.e.,~rises at different $\rho$ and shows significantly different signal strengths. This implies that realistic features, which are not captured in the simulations, play a major role in the experiment. These may be addressed as follows.


\begin{figure}[htp]
    \centering
    \hspace{-7mm}
    \begin{minipage}{1\columnwidth}
        {\includegraphics[height=1.30cm,trim={-0.17cm 8.7cm 0.7cm 1.85cm},clip]{EPS_Sim_colorbar.eps}}
    \end{minipage}\\
    \vspace{1.5mm}
    \begin{minipage}{.75\columnwidth}
        \hspace{2.3mm}\subfloat[\label{subfig:Sim_SF_300fs_s}]{\hspace{-1cm}\includegraphics[height=3.42cm,trim={0cm 1.75cm 7.1cm 1.7cm},clip]{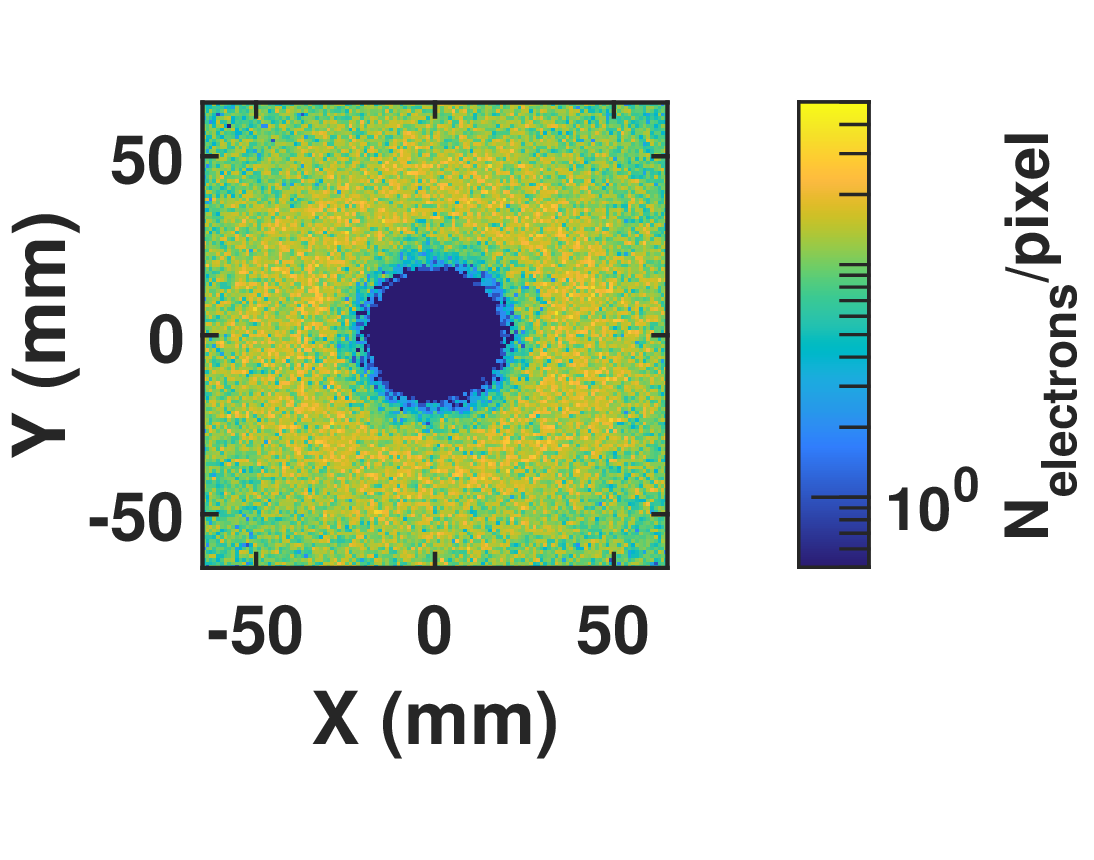}}
        \hspace{1.2mm}\subfloat[\label{subfig:Sim_SF_300fs_w}]{\hspace{0cm}\includegraphics[height=3.42cm,trim={3.2cm 1.75cm 7.1cm 1.7cm},clip]{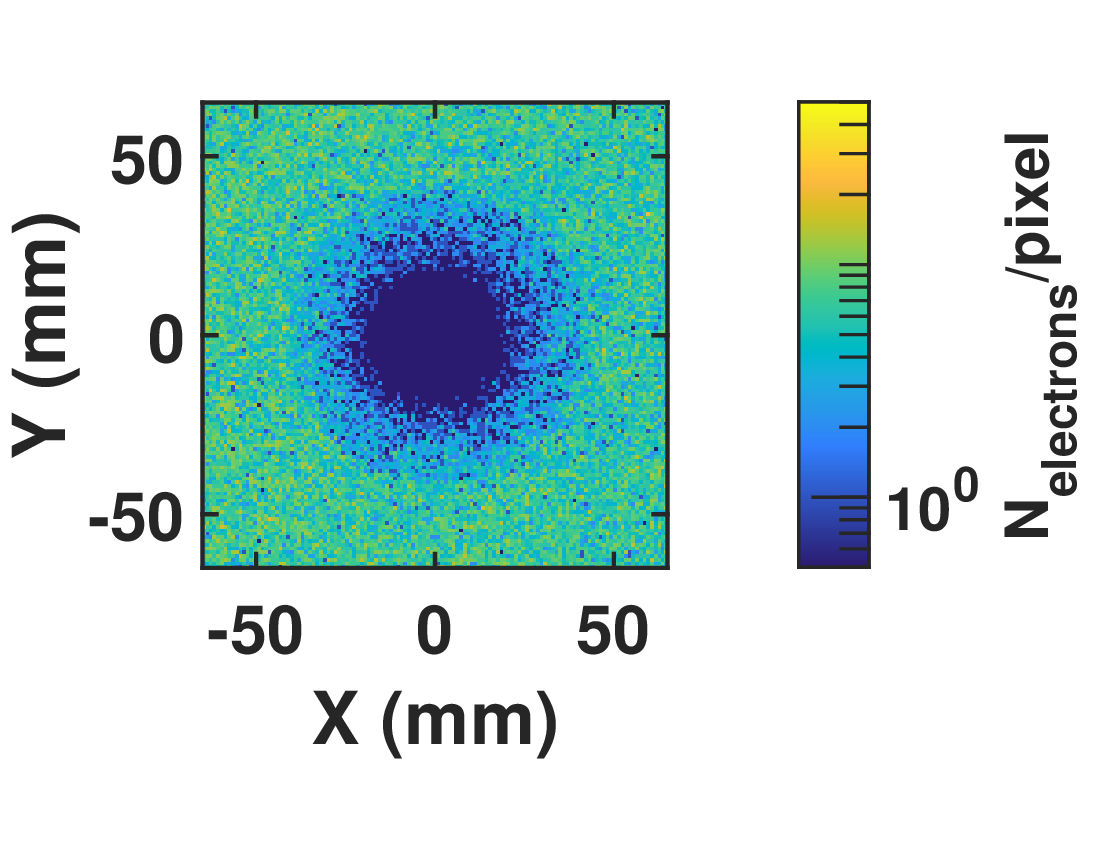}} 
    \end{minipage}\\
    \vspace{-3.2mm}
    \subfloat[\label{subfig:Sim_SF_300fs_lineouts}]{\includegraphics[width=0.8\linewidth]{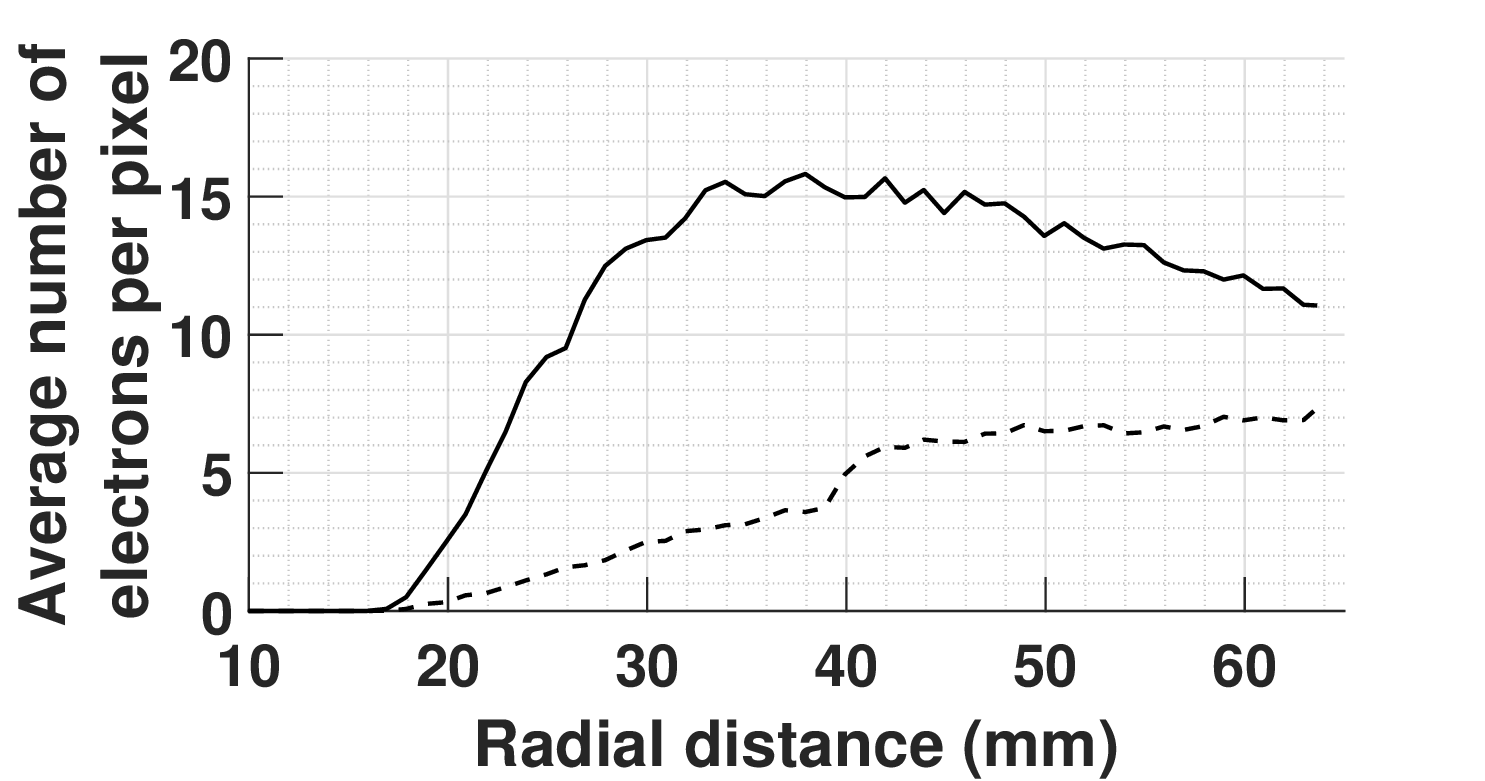}}\\
\caption{Spatial distribution of ejected electrons simulated when the (a) strong pulse and (b) weak pulse leads the other by $300\ \mathrm{fs}$ with randomized relative spatial separation of $\leq2w_0$ between the laser axes. Distributions simulated 30 mm after the focus on a plane perpendicular to the laser axis passing through the origin. The color axis denotes the number of electrons in each $1\ \mathrm{mm} \times 1\ \mathrm{mm}$ pixel across the plane. The corresponding average number of electrons for varying distance from the laser axis is shown by the solid and dashed lines respectively in (c). Simulation details in Appendix \ref{sec:appendix_simulations}.}
\label{fig:300fs_sim_spatial_fluctuation}
\end{figure}

If the peak intensity of the strong pulse delivered in the experiment is lower than that used in the simulations, $\theta_c$ will be larger in the experiment than the simulation i.e.~$N_e$ will begin to rise at larger $\rho$ in the experiment (as $\rho= d\tan \theta$), which can partly explain (i). Any relative spatial fluctuation of the focal volumes of the two beams leading to reduced overlap from shot to shot makes the observed electron distributions less distinct, which can partly explain (ii). This is because each pulse may interact with a volume of gas that is not affected by the other. We show this using simulations that model a randomized relative spatial fluctuation (see Appendix \ref{sec:appendix_simulations}). We show the corresponding distributions when the focal volumes fluctuate relative to each other by applying a relative separation of $\delta \rho \leq 2w_0 $ between the two laser axes per iteration in Figs.~\ref{subfig:Sim_SF_300fs_s} and \ref{subfig:Sim_SF_300fs_w}, with their $N_e(\pm300\ \mathrm{fs},\rho)$ in Fig.~\ref{subfig:Sim_SF_300fs_lineouts}. We see that this results in $N_e$ rising at the same $\rho$ associated with that of the strong pulse. This also increases the overall strength of $N_e(300\ \mathrm{fs})$ (weak pulse leading), making it more similar to $N_e(-300\ \mathrm{fs})$ (strong pulse leading). This is because the strong pulse may now liberate and interact with electrons from He atoms that were not double-ionized by the weak pulse, and eject them down to $\theta_c=\theta_{c,s}$ regardless of whether the weak pulse leads.

\begin{figure}[htp]
    \centering
    \begin{minipage}{1\columnwidth}
        \hspace{0mm}{\vspace{1mm}\includegraphics[height=1.40cm,trim={0.5cm 24.1cm 3.9cm 0.43cm},clip]{EPS_Exp_colorbar.eps}}
    \end{minipage}\\
    \hspace{0mm}
    \begin{minipage}{.75\columnwidth}
        \hspace{0.05cm}\subfloat[\label{subfig:Exp_sample_150fs_s}]{\hspace{-0.8cm}\includegraphics[height=3.44cm,trim={19.4cm 2.0cm 8.5cm 1.5cm},clip]{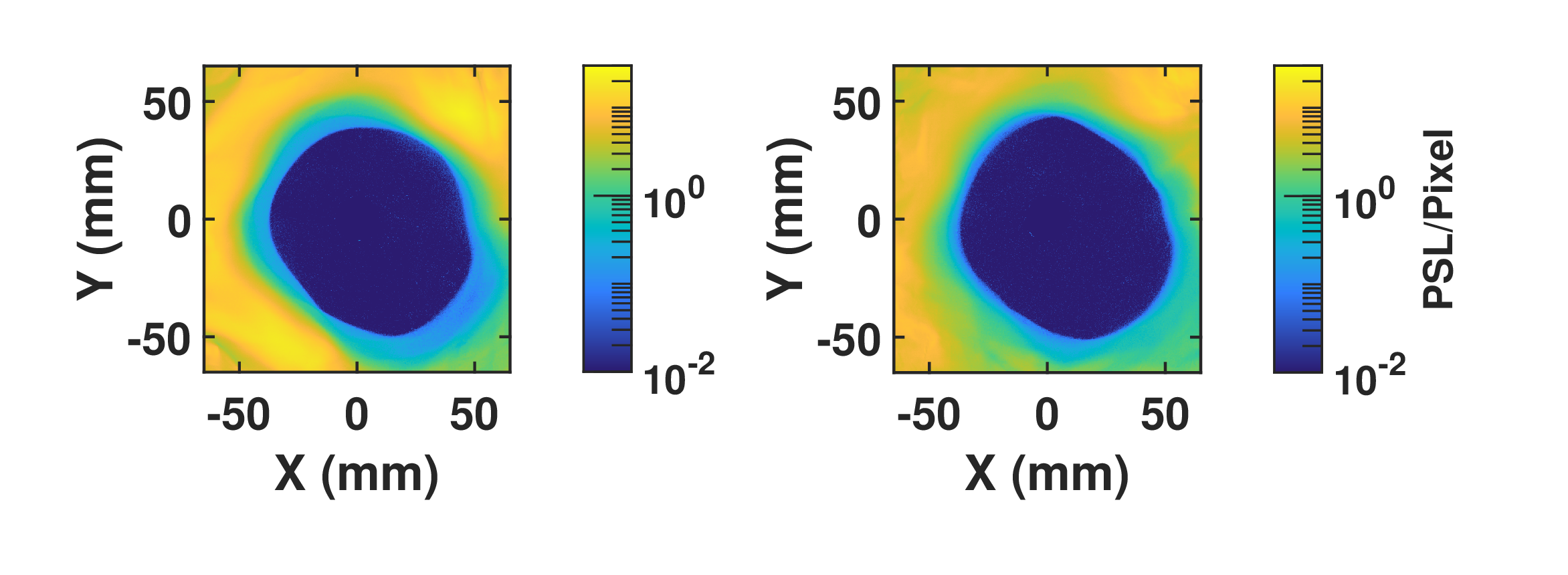}}
        \hspace{-1.0mm}\subfloat[\label{subfig:Exp_sample_150fs_w}]{\hspace{0cm}\includegraphics[height=3.44cm,trim={4.9cm 2.0cm 26.5cm 1.5cm},clip]{EPS_Exp_sample_150fs.eps}} 
    \end{minipage}\\
    \vspace{-3mm}
    \subfloat[\label{subfig:Exp_150fs_lineouts}]{\includegraphics[width=0.8\linewidth]{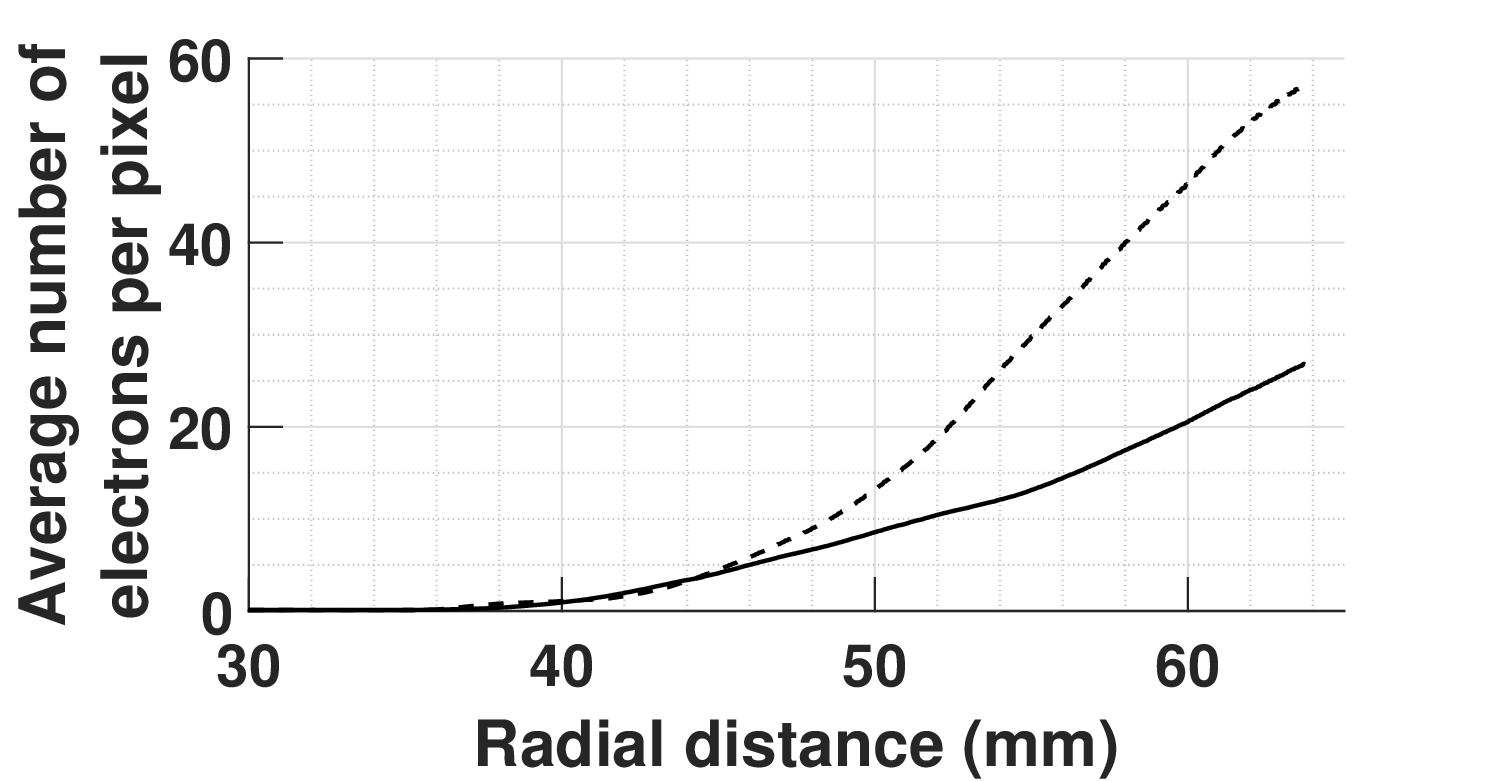}}\\
    \caption{Spatial distribution of ejected electrons measured when the (a) strong pulse and (b) weak pulse leads the other by $150\ \mathrm{fs}$. Distributions measured 30 mm after the focus on a plane perpendicular to the laser axis passing through the origin. The color axis denotes the number of electrons in each $50\ \mathrm{\mu m} \times 50\ \mathrm{\mu m}$ pixel across the plane. The corresponding average number of electrons for varying distance from the laser axis is shown by the solid and dashed lines respectively in (c). Measurement details in Sec.~\ref{sec:methods_section}.}
    \label{fig:150fs_experiment}
    
\end{figure}

Although we observed clearing for $\Delta t=\pm 300 \ \mathrm{fs}$, the differences observed in the experiment and idealized simulations highlight the need to tackle issues such as relative spatial fluctuation between focal volumes. One possible way to address this would be to use different beam geometries that minimize the role of spatial misalignment, such as exploiting the extended ionization volume at larger distances from the focal plane or using a larger clearing beam of commensurate laser intensity. Further, it is important to know how long it takes to clear electrons from the focal volume of interest. To investigate electron clearing for shorter time delays, we performed measurements with $\Delta t = \pm 150 \ \mathrm{fs}$ and present corresponding image plate data in Figs.~\ref{subfig:Exp_sample_150fs_s} and \ref{subfig:Exp_sample_150fs_w} when the strong pulse and weak pulse leads the other, respectively.

\begin{figure}[htp]
    \centering
    \begin{minipage}{1\columnwidth}
        \hspace{0mm}{\vspace{-1mm}\includegraphics[height=1.30cm,trim={0.47cm 8.7cm 0.2cm 1.85cm},clip]{EPS_Sim_colorbar.eps}}
    \end{minipage}\\
    \vspace{2.5mm}
    \hspace{5mm}
    \begin{minipage}{.75\columnwidth}
        \hspace{-0.1cm}\subfloat[\label{subfig:Sim_150fs_s}]{\hspace{-1.05cm}\includegraphics[height=2.55cm,trim={0cm 4.6cm 7.1cm 1.7cm},clip]{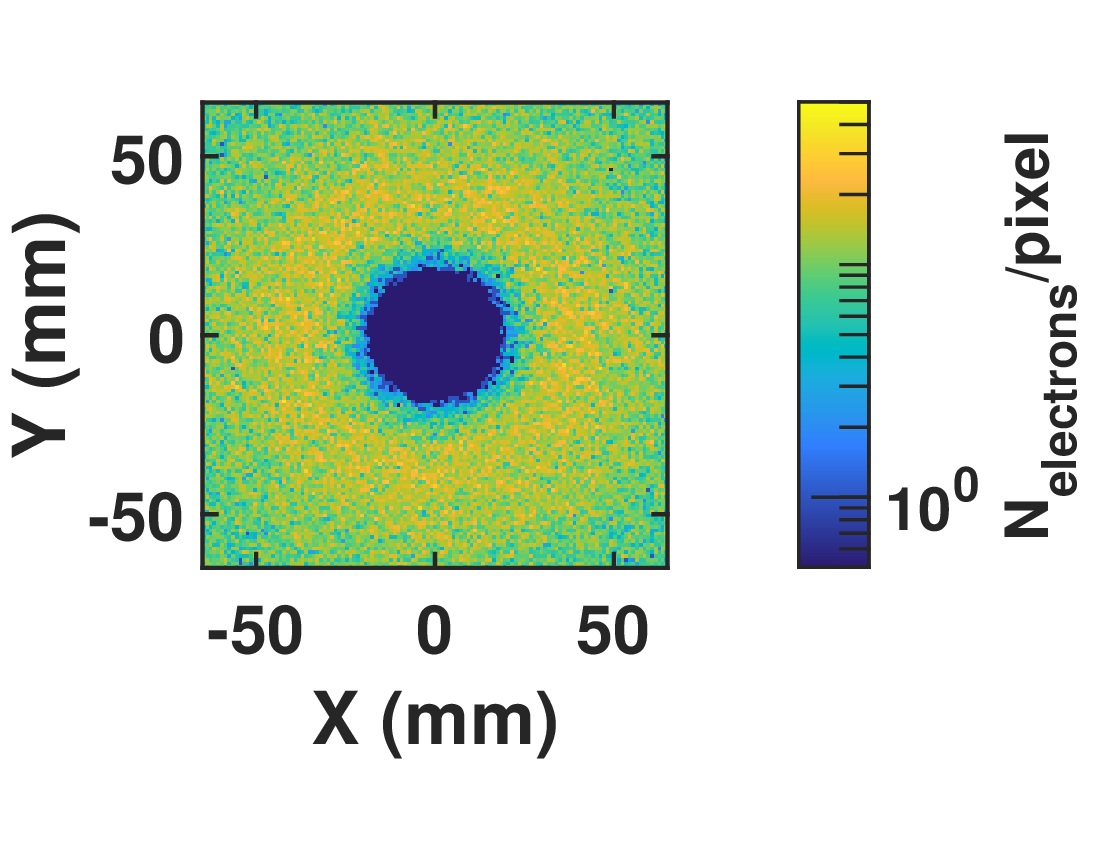}}
        \hspace{1.2mm}\subfloat[\label{subfig:Sim_150fs_w}]{\hspace{0cm}\includegraphics[height=2.55cm,trim={3.2cm 4.6cm 7.1cm 1.7cm},clip]{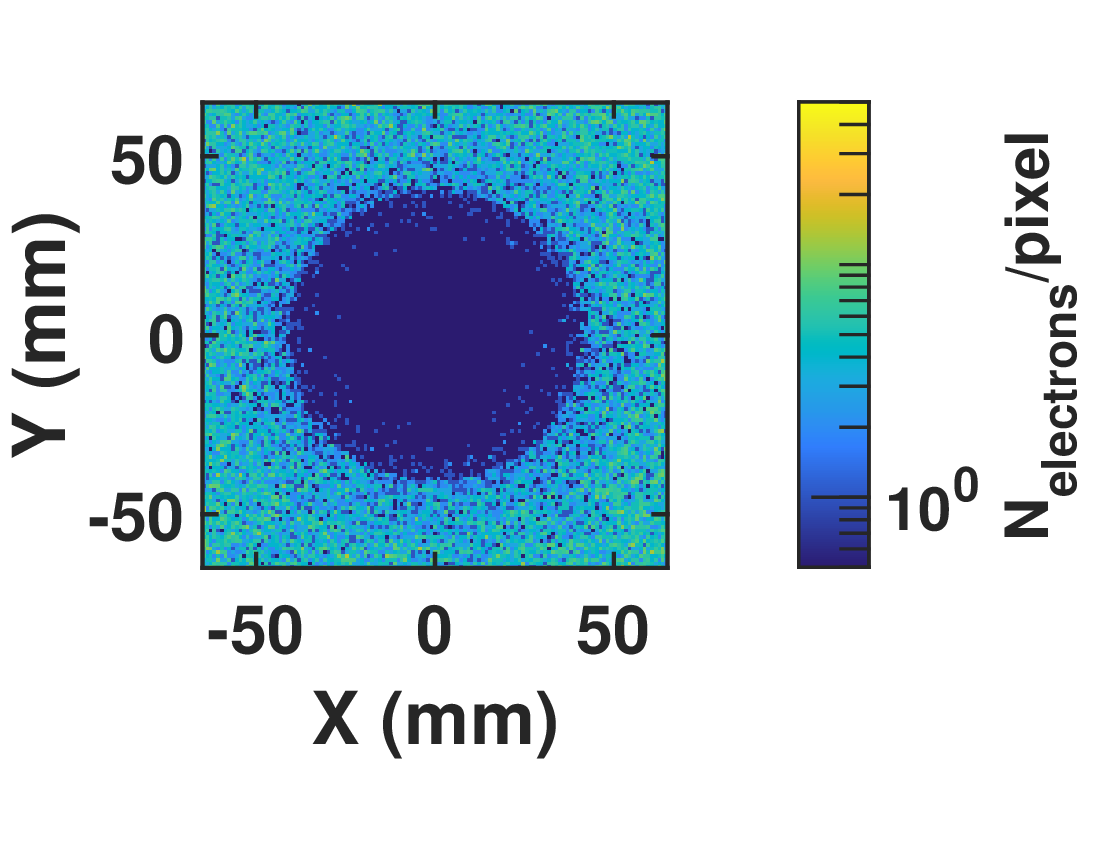}}
    \end{minipage}\\  
    \vspace{2mm}
    \hspace{5mm}
    \begin{minipage}{.75\columnwidth}
        \hspace{-0.1cm}\subfloat[\label{subfig:Sim_L0P1_150fs_s}]{\hspace{-1.05cm}\includegraphics[height=3.42cm,trim={0cm 1.75cm 7.1cm 1.7cm},clip]{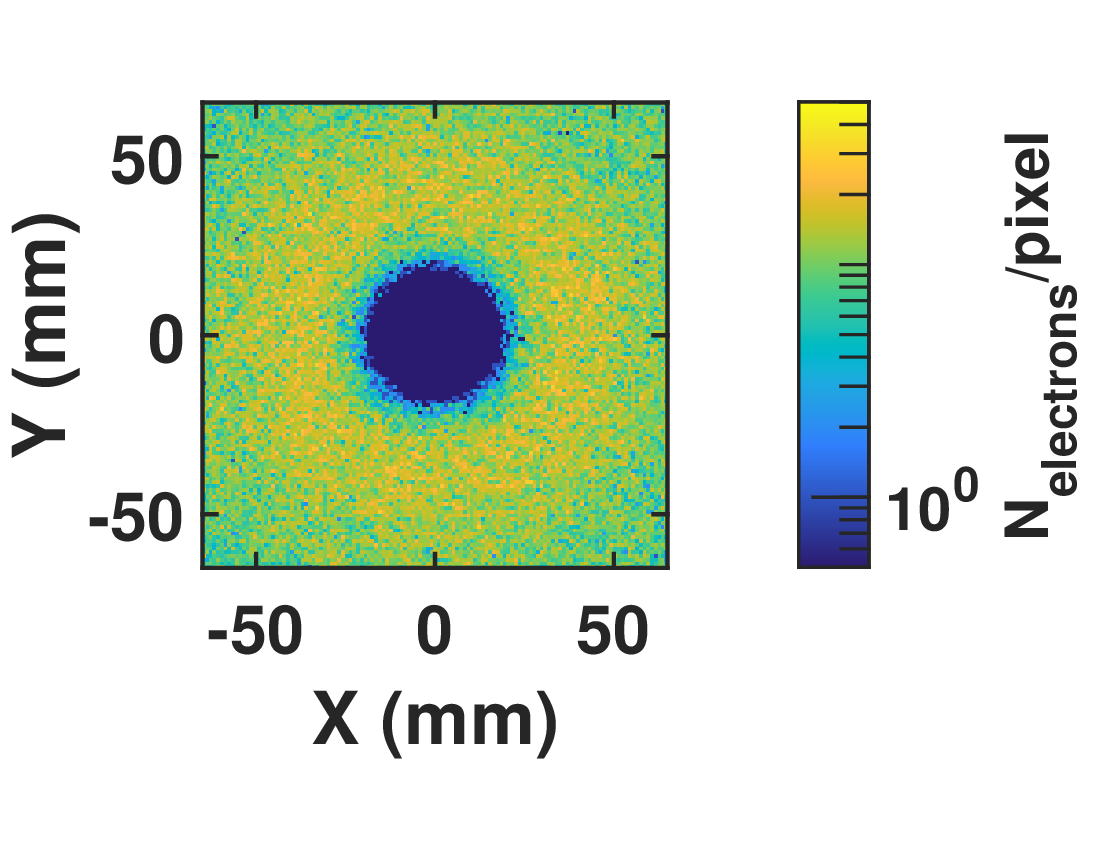}}
        \hspace{1.2mm}\subfloat[\label{subfig:Sim_L0P1_150fs_w}]{\hspace{0cm}\includegraphics[height=3.42cm,trim={3.2cm 1.75cm 7.1cm 1.7cm},clip]{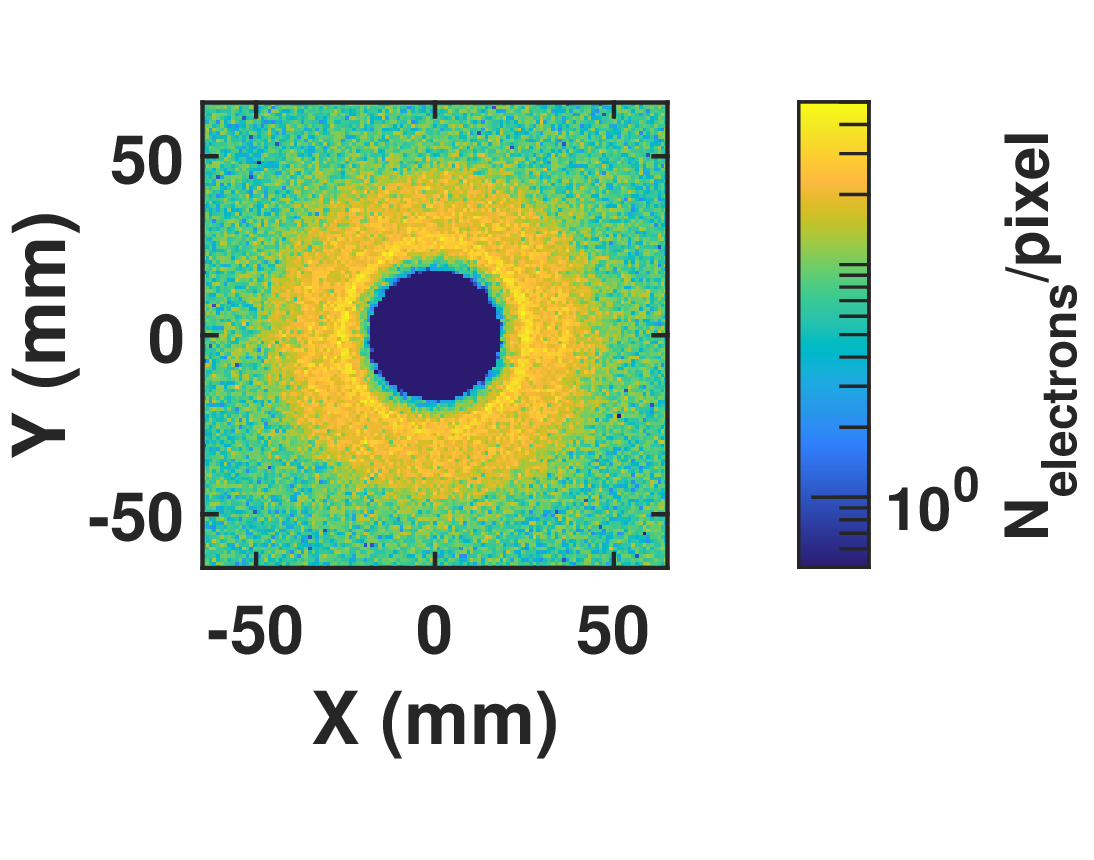}}
    \end{minipage}\\  
    \vspace{1mm}
    \begin{minipage}{1\columnwidth}
        \hspace{8mm}{\vspace{0mm}\includegraphics[height=1.45cm,trim={4.3cm 9.7cm 5.2cm 0.4cm},clip]{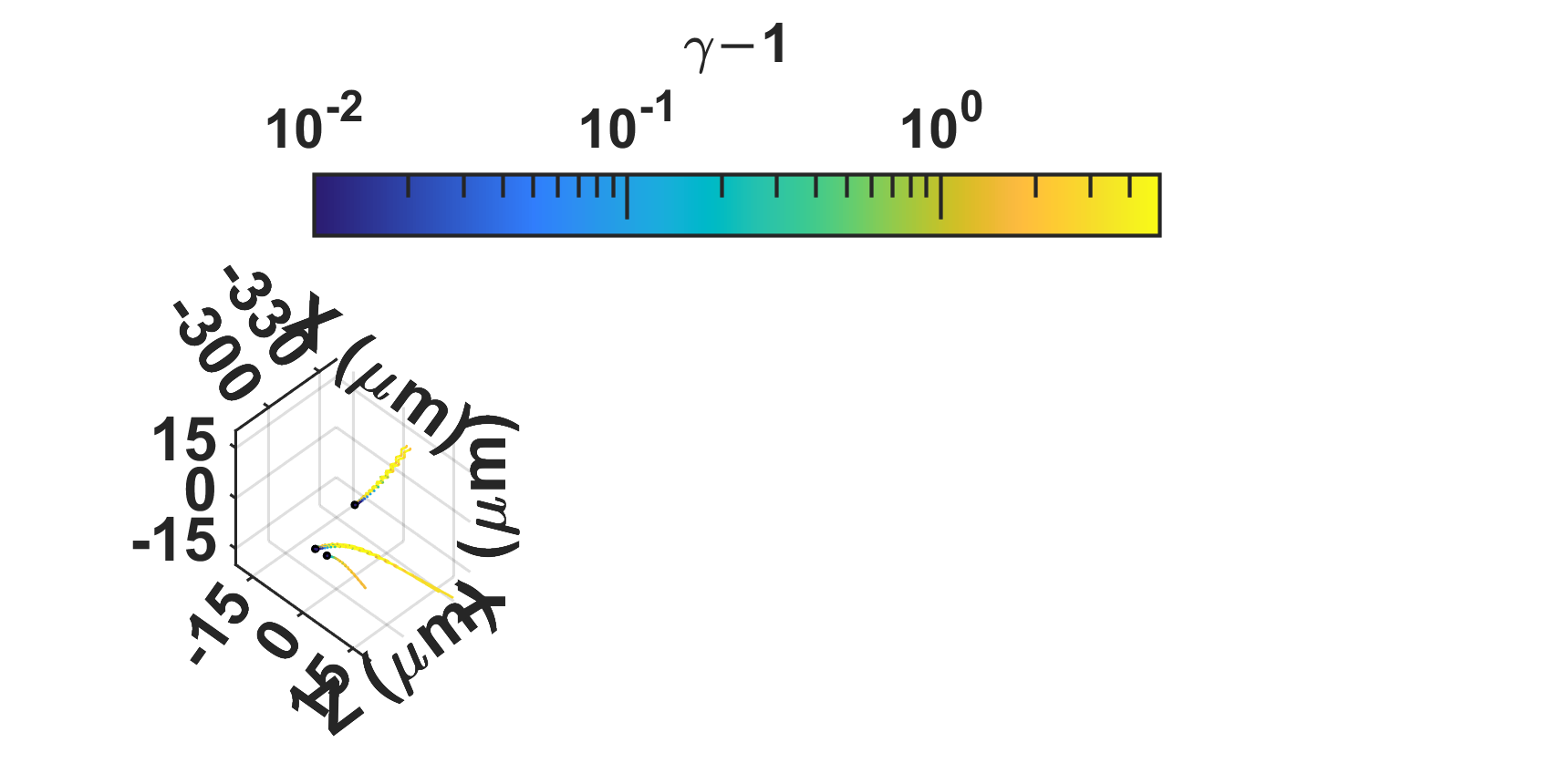}}
    \end{minipage}\\
    \hspace{7mm}
    \begin{minipage}{.75\columnwidth}
        \hspace{0.35cm}\subfloat[\label{subfig:Sim_L0P1_150fs_s_traj}]{\hspace{-1.1cm}\includegraphics[height=3.55cm,trim={6cm 1.9cm 15.5cm 1.7cm},clip]{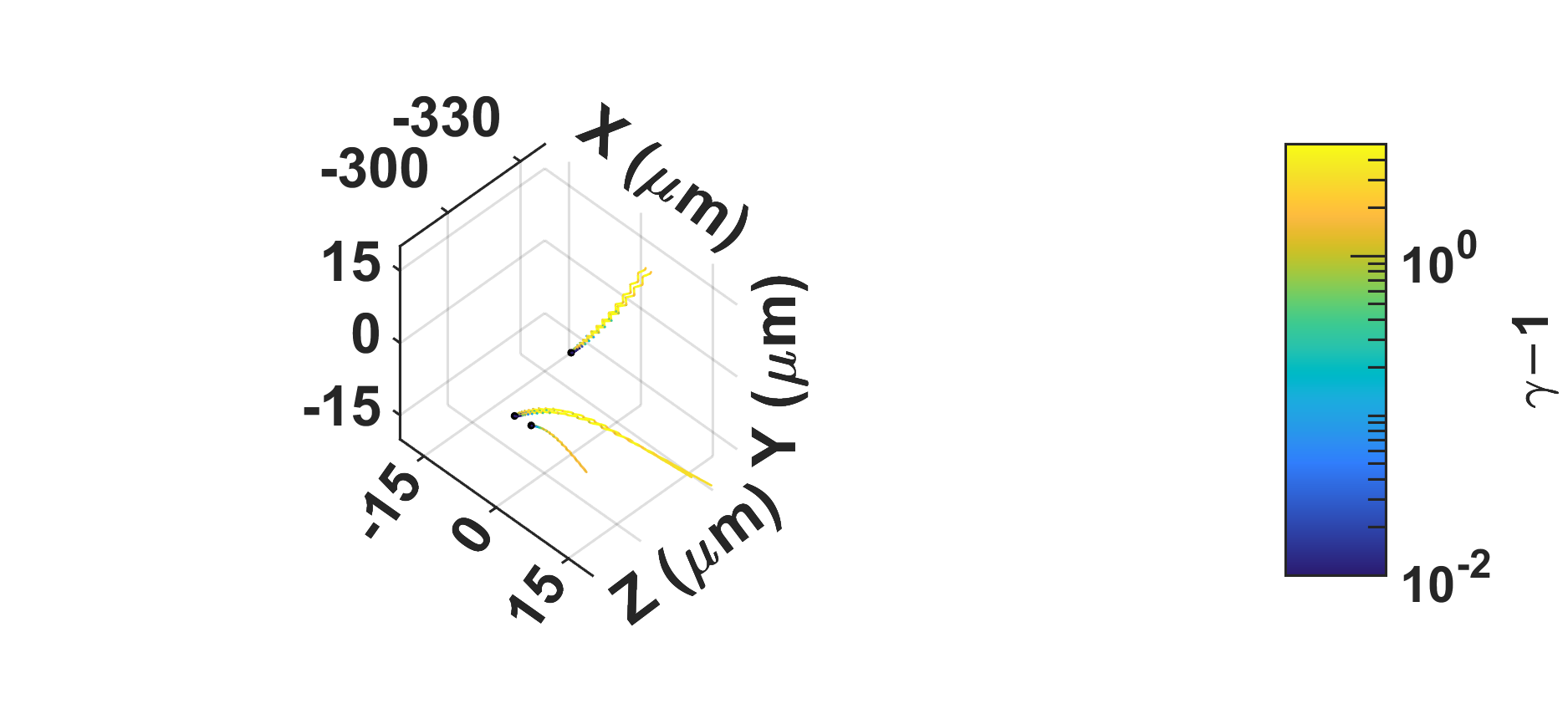}}
        \hspace{-2mm}\subfloat[\label{subfig:Sim_L0P1_150fs_w_traj}]{\hspace{0.4cm}\includegraphics[height=3.55cm,trim={6cm 1.9cm 15.5cm 1.7cm},clip]{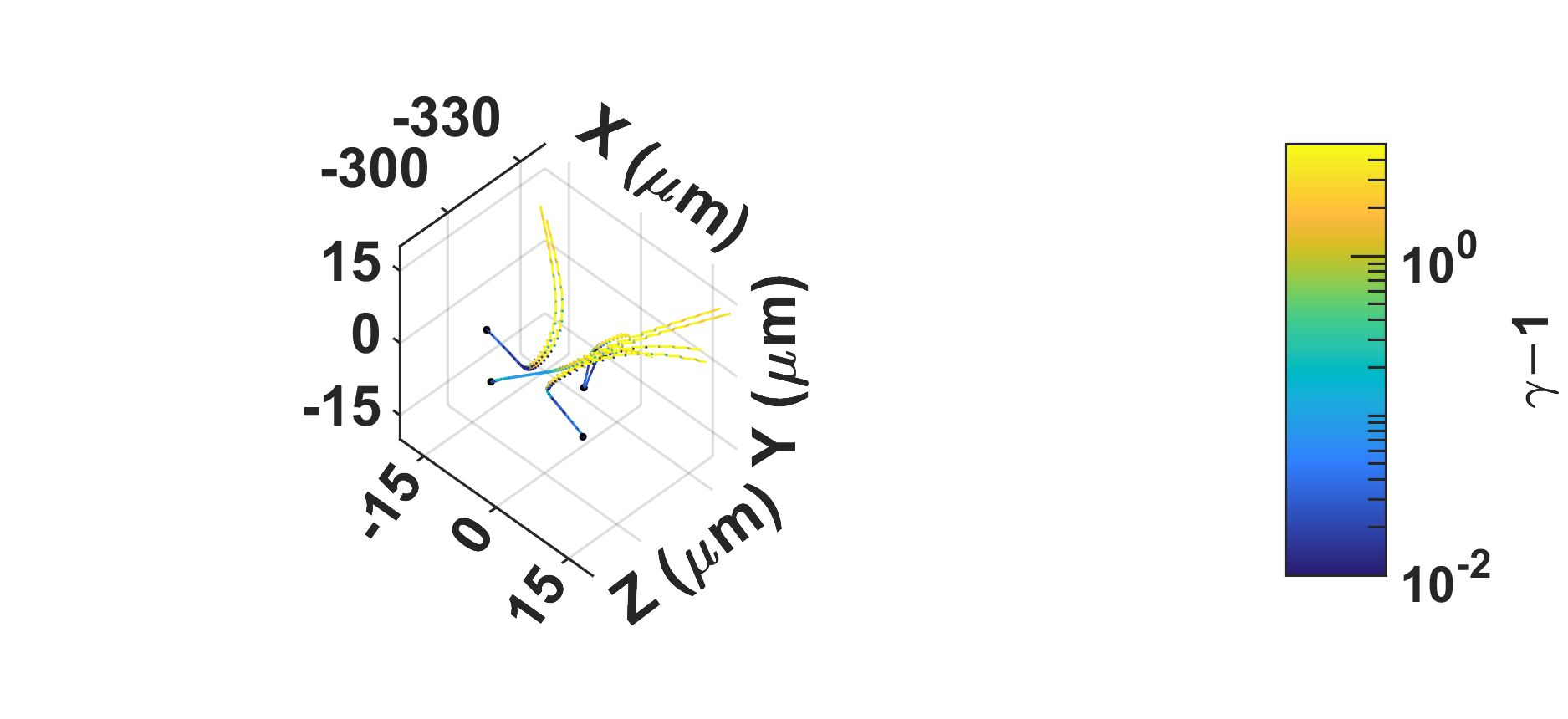}} 
    \end{minipage}
    \caption{Spatial distribution of ejected electrons simulated with $l=0,\ p=0$ Laguerre-Gauss mode beams with (a) strong pulse leading and (b) weak pulse leading the other by 150 fs. Corresponding distributions when the weak pulse is replaced with an $l=0,\ p=1$ mode in (c) and (d) respectively. Color axis denotes number of electrons deposited in each $1\ \mathrm{mm} \times 1 \ \mathrm{mm}$ pixel. Electron trajectories for (b) and (d) shown in (e) and (f) respectively. Birth positions denoted by black points. Color axis denotes the instantaneous $\gamma-1$ through their trajectory. Simulation details in Appendix \ref{sec:appendix_simulations}.}
    \label{fig:Sim_L0P1_data}
    
\end{figure}

We see from the corresponding line profiles in Fig.~\ref{subfig:Exp_150fs_lineouts} that the behavior at $|\Delta t|=150\ \mathrm{fs}$ is contrary to that at 300 fs (Fig.~\ref{subfig:Exp_300fs_lineouts}) in that $N_e$ is stronger when the weak pulse leads. This indicates that the passing of the weak pulse through the focal volume increases the number of electrons available for the strong pulse to interact with over a span of 150 fs. This also contradicts behavior predicted by corresponding simulations with an $l=0,\ p=0$ Gaussian laser profile at 150 fs, presented in Figs.~\ref{subfig:Sim_150fs_s} and \ref{subfig:Sim_150fs_w}. However, we see from Fig.~\ref{subfig:Focal_spot_weak} that the spatial profile of the weak pulse has a distinct ring-like feature outside the central spot with sufficient intensity to fully ionize He. Such features are often expected in the focus of a flattop laser pulse. Further, electrons ionized by the secondary maxima outside the central spot could contribute to the overall population within the focal volume of the strong pulse by expelling electrons toward the laser axis. Note that the volume of the ring-like feature is larger than the central spot and thus a large number of electrons can be generated therein. To better understand the dynamics involved, we simulated this interaction replacing the weak pulse with a Laguerre-Gauss $l=0,\ p=1$ of same $a_{0w}$ and $w_0$ as the $l=0,\ p=0$ mode (see Appendix \ref{sec:appendix_simulations}). We used this as an approximation to recreate the ring-like feature outside the central spot in Fig.~\ref{subfig:Focal_spot_weak} while maintaining the characteristic peak intensity of the weak pulse. We show the resultant electron distributions in Figs.~\ref{subfig:Sim_L0P1_150fs_s} and \ref{subfig:Sim_L0P1_150fs_w}.  

It is evident that when the weak pulse with an $l=0,\ p=1$ mode leads the strong pulse (Fig.~\ref{subfig:Sim_L0P1_150fs_w}), the number of electrons ejected onto the observation plane at small angles is higher than that when the strong pulse leads (Fig.~\ref{subfig:Sim_L0P1_150fs_s}). This implies that there is indeed a higher concentration of electrons that remain in the focal volume over 150 fs after the passing of the weak pulse, which are subsequently ejected by the following strong pulse toward small angles. In comparison, such electrons do not appear in the absence of a ring-like feature in the weak pulse, resulting in the lack of electrons being ejected at small angles (Fig.~\ref{subfig:Sim_150fs_w}). This may be further explained using the trajectories of electrons appearing at small angles ($\theta\sim 35^{\circ}$), corresponding to final $\gamma-1\sim4.5$ in Figs.~\ref{subfig:Sim_L0P1_150fs_s} and \ref{subfig:Sim_L0P1_150fs_w}, shown in Figs.~\ref{subfig:Sim_L0P1_150fs_s_traj} and \ref{subfig:Sim_L0P1_150fs_w_traj} respectively. Here, $(\gamma-1)\times0.511$ MeV is the kinetic energy of the electron. We see in Fig.~\ref{subfig:Sim_L0P1_150fs_s_traj} that when the strong pulse leads, electrons are ponderomotively accelerated and ejected radially outward from the focal volume. However, we see in Fig.~\ref{subfig:Sim_L0P1_150fs_w_traj} that when the more realistic weak pulse leads, electrons that are ejected at similar angles are born at larger $\rho$ from the ring-like feature, injected inward with non-relativistic energies ($\gamma-1 \lesssim 0.1$) and then ejected radially outward by the strong pulse with $\gamma-1\sim4.5$. It is straightforward to calculate the drift length for these electrons during the initial 150 fs after ionization using $\gamma=1/\sqrt{1-v^2/c^2}$. Using $\gamma-1\sim0.02$ to 0.1, we get a drift length $\sim 9-18 \ \mathrm{\mu m}$ that is in agreement with Fig.~\ref{subfig:Sim_L0P1_150fs_w_traj}. This implies that the electrons ionized by secondary maxima of the weak pulse can potentially cause a temporary increase in electron density within the focal volume that is counter-productive to ponderomotive clearing. Over the longer time scale of 300 fs, most of these electrons pass through and leave the central focal volume. This highlights that although it is possible to clear the focal volume with ponderomotive clearing, it is less straightforward to achieve efficient clearing, which depends on various features of the clearing beam and could vary for different time delays. This will be explored in greater detail through further experiments and simulations.  

\section{Conclusion}\label{sec:conclusion_section}

We showed using simulation and experiment that it is possible to temporarily clear real electrons from the focal volume of a petawatt laser using a clearing pulse. We observed a reduction in the average number of electrons ejected to $\theta\lesssim63^{\circ}$ (Fig.~\ref{subfig:Exp_300fs_lineouts}) over 300 fs after the clearing pulse, indicating a reduction in the number of electrons within the focal volume. We noted using simulations (Figs.~\ref{subfig:Sim_300fs_lineouts} and \ref{subfig:Sim_SF_300fs_lineouts}) that the effective clearing observed in the experiment depends on the peak intensity of the beams delivered to the focus and the spatial overlap of the focal volumes, which emphasizes the need to address spatial jitter. Although a copropagating beam geometry may provide a relatively high probability of overlapping focal volumes, deploying the two pulses at an angle such as in a perpendicular geometry allows spatial separation of ejected electrons. This could enable a quantitative measurement of the reduction in electron density in the near term. 

We further showed using experiment (Fig.~\ref{fig:150fs_experiment}) and simulations (Fig.~\ref{fig:Sim_L0P1_data}) that secondary maxima outside the central spot in the spatial profile of the clearing pulse can directly affect the concentration of electrons in the focal pulse for $\sim 150\ \mathrm
{fs}$. We investigated the role of a ring-like feature exhibited by the weak pulse used for clearing (Fig.~\ref{subfig:Focal_spot_weak}) by modeling it with a Laguerre-Gauss $l=0, \ p=1$ mode to introduce a secondary maxima at $\rho>0$. We showed that this can cause an injection of electrons from large $\rho$ to small $\rho$ and temporarily increase the electron density in the focal volume. This highlights the need to investigate the optimal range of laser parameters in order to provide efficient clearing of the focal volume for precision-measurement based experiments that require an ideal void of electrons. We further note that the dynamics of electron expulsion may provide insight for future experiments at $\sim 10^{24} \ \mathrm{W/cm^2}$ and beyond where protons are expected to be ponderomotively expelled from the focus, resulting in a near-ideal vacuum.

\begin{acknowledgments}

The authors acknowledge the University of Maryland supercomputing resources (http://hpcc.umd.edu) made available for conducting the research reported in this paper. The authors would like to thank Lester Putnam and Matias Calderon for technical help in preparation for the experiment and everyone at CLPU for their conducive involvement with the experiment. This work was supported by the National Science Foundation (Grant No.~PHY2010392, PHY2308905, PHY2329970), Natural Sciences and Engineering Research Council of Canada (Grant No.~ RGPIN-2019-05013), Ministerio de Ciencia e Innovaci\'on of Spain (PID2022-140593NB-C22).
\end{acknowledgments}

\appendix

\section{Description of simulations}\label{sec:appendix_simulations}

As the gas density is maintained sufficiently low ($\lesssim 10^{13}$ particles$/\mathrm{cm^3}$) to avoid collective effects, we simulate this interaction by numerically solving the relativistic Lorentz force equation for an electron in the laser field. This is given by $d(\gamma m\vec{v})/dt = -e(\vec{E}_{tot} +\vec{v}\times\vec{B}_{tot} )$, where $\vec{E}_{tot} = \vec{E}_s + \vec{E}_w$ and $\vec{B}_{tot}  = \vec{B}_s + \vec{B}_w$ are the total electric and magnetic fields in the presence of the strong and weak laser pulses whose fields are ($\vec{E}_s$, $\vec{B}_s$) and ($\vec{E}_w$, $\vec{B}_w$) respectively. We neglect radiation reaction for the range of laser intensities involved herein. We use the description of field components provided by W.L.~Erikson and Surendra Singh\cite{erikson_polarization_1994}, which we consider a good fit for modeling paraxial beams \cite{peatross_vector_2017} such as the f/10 focus used in our experiment. The field at some point $(x,y,z)$ at time $t$ due to a single laser pulse with an $l=0,\ p=0$ Laguerre-Gauss spatial profile polarized along $\hat{x}$ and propagating along $\hat{z}$ with the focal plane at $z=0$ is\cite{longman_modeling_2022}

\begin{equation} \label{eqn:E_field}
    \vec{E}(x,y,z,t) = E_0\left[ \hat{x} + \frac{xy}{2Z_i^2}\hat{y} - \frac{ix}{Z_i} \hat{z} \right]\psi(x,y,z) f(t)
\end{equation}

\begin{equation} \label{eqn:B_field}
    \vec{B}(x,y,z,t) = \frac{E_0}{c}\left[ \frac{xy}{2Z_i^2}\hat{x} + \hat{y} - \frac{iy}{Z_i}\hat{z} \right]\psi(x,y,z) f(t)
\end{equation}

\noindent where $E_0$ is the peak electric field strength, $Z_i$ is the complex beam parameter $z_R + iz$ for Rayleigh length $z_R$, $\psi(x,y,z)$ is the $l=0,\ p=0$ spatial profile and $f(t)$ is the temporal profile. For reference, $\psi(x,y,z) = (w_0/w(z)) \mathrm{exp}( -k\rho^2/2Z_i) \mathrm{exp}(-i\tan^{-1}(z/z_R))$ and $f(t) = \mathrm{exp}(-ln(2)(kz-\omega_0t)^2/(2\omega_0^2\tau^2)+i(kz-\omega_0t))$. Here, $w(z) = w_0\sqrt{1+z^2/z_R^2}$, $\rho = \sqrt{x^2+y^2}$, $\omega_0 = ck$, $k = 2\pi/\lambda_0$, $c$, $\tau$ are the beam radius, radial distance, angular frequency of light, magnitude of the wave vector, $\vec{k}$, speed of light in vacuum and the full-width-at-half-maximum pulse duration. Here, we use $\lambda_0 = 0.8 \ \mathrm{\mu m}$, $w_0 = 10 \ \mathrm{\mu m}$ and $\tau=40 \ \mathrm{fs}$. The total field at any point is given as $\vec{E}_{tot} = \vec{E}_s + \vec{E}_w$ and $\vec{B}_{tot}  = \vec{B}_s + \vec{B}_w$, where ($\vec{E}_s$, $\vec{B}_s$) and ($\vec{E}_w$, $\vec{B}_w$) are the fields of the strong and weak pulse respectively. Both pulses have the same polarization and may be described for each type of simulation as follows.

To describe two pulses with perfect overlap of focal volumes, $\vec{E}_s = \vec{E}$ and $\vec{B}_s = \vec{B}$ for the strong pulse, and $\vec{E}_w(t) = (E_{0w}/E_{0})\vec{E}(t+\Delta t)$ and $\vec{B}_w(t) = (E_{0w}/E_{0})\vec{B}(t+\Delta t)$ for the weak pulse with peak electric field strength $E_{0w}$. The electric field strengths may also be written as $E_0 = \sqrt{2I_{0s}/(c\epsilon_0)}$ and $E_{0w} = \sqrt{2I_{0w}/(c\epsilon_0)}$, where $\epsilon_0$ is the permittivity of free space. For all simulations presented in this manuscript, we used $I_{0s}=5\times10^{19}\ \mathrm{W/cm^2}$, $I_{0w}=1.25\times10^{19}\ \mathrm{W/cm^2}$ and assumed the same beam waist $w_0=10\ \mathrm{\mu m}$ for both beams for simplicity.

To include relative spatial separation between the two laser axes on each shot, we applied a shift of coordinates to the weak pulse prior to calculating the total field. The shift along the $\vec{x}$ and $\vec{y}$ axes are given by $\delta x$ and $\delta y$, which vary randomly for different iterations of the simulation such that $\sqrt{(\delta x)^2+(\delta y)^2}\leq 2w_0$ to give $\vec{E}_w(t) = (E_{0w}/E_0)\vec{E}(x+\delta x, y+\delta y, t+\Delta t)$ and $\vec{B}_w(t) = (E_{0w}/E_0)\vec{B}(x+\delta x, y+\delta y, t+\Delta t)$. This models a weak pulse whose center is separated from the center of the strong pulse at most by $2w_0$ at arbitrary $\phi$ along the transverse plane on each shot. 

For simulations where the weak pulse is replaced with an $l=0,\ p=1$ mode, the corresponding fields are written as \cite{longman_modeling_2022}

\begin{equation} \label{eqn:E_field_l0p1}
    \vec{E}_w = E_{0w}\left[ \hat{x} + \frac{xy(1+4L_1)}{2Z_i^2}\hat{y} - \frac{ix(1+2L_1)}{Z_i} \hat{z} \right]\psi^0_1f(t+\Delta t)
\end{equation}

\begin{equation} \label{eqn:B_field_l0p1}
    \vec{B}_w = \frac{E_{0w}}{c}\left[ \frac{xy(1+4L_1)}{2Z_i^2}\hat{x} + \hat{y} - \frac{iy(1+2L_1)}{Z_i}\hat{z} \right]\psi^0_1f(t+\Delta t)
\end{equation}

\noindent where $L_1 = (z_R/Z_i^*)/(1-2\rho^2/w(z)^2)$ and $\psi^0_1$ is the $l=0,\ p=1$ spatial profile given by $\psi^0_1 =(w_0/w(z))(1-2\rho^2/w(z)^2)\mathrm{exp}(-k\rho^2/(2Z_i))\mathrm{exp}(-3i\tan^{-1}(z/z_R))$.

To solve the equation of motion for each electron we need to find its birth time and position. We achieve this by starting with a random population of gaseous He atoms in a box that is sufficiently sized to include all electrons of interest that impinge on the image plate. We use such an initial box with boundaries defined by $(x_B,y_B,z_B) = (\pm12w_0,\pm12w_0,\pm4z_R)$. The particles are stationary as the motion of neutral atoms and ions is negligible over durations $\ll$ 1 ns for laser intensities $\ll 10^{24}\ \mathrm{W/cm^2}$. We assume barrier-suppression ionization (BSI) \cite{augst_laser_1991}, which gives an ionization intensity threshold $I_{th}\ (\mathrm{W/cm^2}) = 4\times 10^9 U_{ion}^4(\mathrm{eV})/Z^2$, where $U_{ion}$ is the ionization potential corresponding to a specific ionization stage of an atom \cite{noauthor_nist_nodate}. We use a simplistic approach to smooth the onset of BSI by smoothing the barrier over $0.5\times$ to $2\times I_{th}$. Specifically, each particle at some position $(x,y,z)$ creates an electron at the same location at some instant, $t=t_0$, when $I_{tot}(x,y,z,t_0) \geq \delta I_{th}$, where $I_{tot}(x,y,z,t_0)=c\epsilon_0|\vec{E}_{tot}(x,y,z,t_0)|^2/2$ is the instantaneous intensity of the total field and $\delta$ is a random number $\in (0.5,2.0)$. The accuracy of ionization rates around $I_{th}$ are not critical for the purpose of this calculation as $I_{0,tot}\gg I_{th}$ for helium $(<10^{16} \mathrm{W/cm^2})$, where $I_{0,tot}$ is the peak intensity of the total field ($\gtrsim10^{19} \ \mathrm{W/cm^2}$). While it may have an effect when $I_{th}\sim I_{0,tot}$ and in such cases it may be useful to adopt a more sophisticated approach like the Ammosov–Delone–Krainov (ADK) theory \cite{ammosov_tunnel_1986}, we note that the bulk of electrons observed using the image plate are ionized at $I_{th} \ll I_{0,tot}$. Once the electrons are ionized we numerically solve their equations of motion in free space up to an instant when the trailing pulse reaches $z=4z_R+100c\tau$ to allow all electrons sufficient time to reach steady state. 

Due to the excessive computational resource required to model the interaction at scale with the experiment, we performed our simulations on a reasonable scale to achieve sufficient sampling to compare with the experiment. We simulated each set of parameters using $10^{-8} \ \mathrm{mbar}$ of He gas for 500 iterations (equivalent to a single shot with $5\times 10^{-6}\ \mathrm{mbar}$). This allowed reasonable sampling with pixel size $1\ \mathrm{mm} \times1\ \mathrm{mm}$ on a plane 30 mm after the laser focus.

\section{Converting PSL values to number of electrons}\label{sec:appendix_PSL_to_electrons}

For an image plate measurement with some time delay $\Delta t$ between the pulses, we define $g(\Delta t, \rho,\phi)$ as the PSL value stored in a pixel located at some ($\rho,\phi$) on the image plate. We account for the acute incidence angle of electrons on the image plate that results in a larger PSL deposited relative to that under normal incidence by a factor of $1/\cos\theta$, where $\theta = \tan^{-1}(\rho/d)$. Under paraxial conditions, we assume\cite{moore_observation_1995} $\gamma-1 = 2/\tan^2 \theta $ to estimate the average energy lost by an electron with kinetic energy $(\gamma(\rho)-1)m_ec^2$ passing through an Al filter of thickness $12 \ \mathrm{\mu m}/\cos\theta$. We use the continuous slowing down approximation range for Al \cite{noauthor_stopping_nodate,noauthor_nist_nodate-1} and define the average loss in kinetic energy as $L_{KE}(\rho)$. We then find the PSL value that a single electron with an effective kinetic energy $\approx (\gamma(\rho)-1)m_ec^2 - L_{KE}(\rho)$ would deposit on average in a BAS-MS image plate, based on calibrations performed by G.~Boutoux et al.\cite{boutoux_study_2015} and call it $h(\rho,\phi)$. The number of electrons thus incident on a pixel at given $\rho$ and $\phi$ may be estimated as $n_e(\Delta t,\rho,\phi) = g(\Delta t,\rho,\phi)\sqrt{1+\rho^2/d^2}/h(\rho,\phi)$.

\section*{References}

\bibliography{Bib_Research_Articles,Bib_Books,Bib_Webpages}

\end{document}